\documentclass[letterpaper,twocolumn,10pt]{article}

\PassOptionsToPackage{hyphens}{url}

\usepackage{usenix2019_v3}

\usepackage{amsmath,amssymb}
\usepackage{todonotes}
\usepackage{tikz}
\usetikzlibrary{positioning,shadows,fit,calc,shapes}
\usepackage{hyperref}
\usepackage{booktabs}
\usepackage{tabularx}
\usepackage{array}
\usepackage{soul}
\usepackage{algorithm}
\usepackage[noend]{algpseudocode}
\usepackage[inline]{enumitem}
\usepackage{graphicx}
\usepackage{multirow}
\usepackage{xcolor}
\usepackage{fontawesome}
\usepackage{fancyvrb}
\usepackage[font={small}]{caption, subcaption}
\usepackage[official]{eurosym}

\hypersetup{
	linkcolor=black
}

\newcommand{\ie}{i.e.}
\newcommand{\eg}{e.g.}
\newcommand{\etal}{\emph{et al.}}

\newcommand{\rda}{related-domain attacker}

\newcommand{\capab}[1]{\texttt{#1}}
\newcommand{\domain}[1]{\texttt{#1}}

\definecolor{darkred}{RGB}{190,0,0}
\definecolor{darkgreen}{RGB}{0,140,0}

\newcommand\safe{\textcolor{cyan}{\faCheck}}
\newcommand\vuln{\textcolor{darkred}{\faExclamationCircle}}
\newcommand\notapply{\textcolor{gray}{\faMinus}}
\newcommand\boh{\notapply}

\newcommand{\dataVDomains}{887}
\newcommand{\dataVSubdomains}{1,520}

\newcommand{\dataPDomains}{13,911}
\newcommand{\dataPSubdomains}{189,017}

\newcommand{\dataVdDomains}{201}
\newcommand{\dataVdSubdomains}{260}

\newcommand{\dataVsDomains}{699}
\newcommand{\dataVsSubdomains}{1,260}

\newcommand{\dataVcDomains}{13,532}
\newcommand{\dataVcSubdomains}{187,498}
\newcommand{\dataCloudNProviders}{6}

\newcommand{\dataPortscanAll}{148}
\newcommand{\dataPortscanTCP}{128}
\newcommand{\dataPortscanUDP}{20}

\newcommand{\dataTotThirdPartyServices}{26}
\newcommand{\dataTotDynDNS}{5}
\newcommand{\dataTotServices}{31}

\newcommand{\dataEduTotSites}{1,229}
\newcommand{\dataEduAvgSubdomains}{6,033}

\newcommand{\dataEduVulnPercentage}{7.32\%}

\newcommand{\dataComVulnPercentage}{1.81\%}

\newcommand{\dataOriginsCSPScript}{1,144}
\newcommand{\dataDomainsCSPScript}{260}
\newcommand{\dataOriginsCSPScriptVulnerableWeb}{901}
\newcommand{\dataDomainsCSPScriptVulnerableWeb}{212}
\newcommand{\dataOriginsCSPScriptVulnerableReldom}{0}
\newcommand{\dataDomainsCSPScriptVulnerableReldom}{0}
\newcommand{\dataOriginsCSPScriptPotentiallyVulnerableReldom}{9}
\newcommand{\dataDomainsCSPScriptPotentiallyVulnerableReldom}{8}

\newcommand{\dataOriginsCSPStyle}{961}
\newcommand{\dataDomainsCSPStyle}{232}
\newcommand{\dataOriginsCSPStyleVulnerableWeb}{930}
\newcommand{\dataDomainsCSPStyleVulnerableWeb}{225}
\newcommand{\dataOriginsCSPStyleVulnerableReldom}{0}
\newcommand{\dataDomainsCSPStyleVulnerableReldom}{0}
\newcommand{\dataOriginsCSPStylePotentiallyVulnerableReldom}{2}
\newcommand{\dataDomainsCSPStylePotentiallyVulnerableReldom}{2}

\newcommand{\dataOriginsCSPObject}{1,027}
\newcommand{\dataDomainsCSPObject}{250}
\newcommand{\dataOriginsCSPObjectVulnerableWeb}{586}
\newcommand{\dataDomainsCSPObjectVulnerableWeb}{118}
\newcommand{\dataOriginsCSPObjectVulnerableReldom}{12}
\newcommand{\dataDomainsCSPObjectVulnerableReldom}{7}
\newcommand{\dataOriginsCSPObjectPotentiallyVulnerableReldom}{34}
\newcommand{\dataDomainsCSPObjectPotentiallyVulnerableReldom}{23}

\newcommand{\dataOriginsCSPFrame}{967}
\newcommand{\dataDomainsCSPFrame}{229}
\newcommand{\dataOriginsCSPFrameVulnerableWeb}{619}
\newcommand{\dataDomainsCSPFrameVulnerableWeb}{140}
\newcommand{\dataOriginsCSPFrameVulnerableReldom}{45}
\newcommand{\dataDomainsCSPFrameVulnerableReldom}{20}
\newcommand{\dataOriginsCSPFramePotentiallyVulnerableReldom}{129}
\newcommand{\dataDomainsCSPFramePotentiallyVulnerableReldom}{43}

\newcommand{\dataOriginsCSPFraming}{1,676}
\newcommand{\dataDomainsCSPFraming}{360}
\newcommand{\dataOriginsCSPFramingVulnerableWeb}{247}
\newcommand{\dataDomainsCSPFramingVulnerableWeb}{28}
\newcommand{\dataOriginsCSPFramingVulnerableReldom}{97}
\newcommand{\dataDomainsCSPFramingVulnerableReldom}{32}
\newcommand{\dataOriginsCSPFramingPotentiallyVulnerableReldom}{302}
\newcommand{\dataDomainsCSPFramingPotentiallyVulnerableReldom}{74}

\newcommand{\dataOriginsCORSVulnerableWeb}{2,030}
\newcommand{\dataDomainsCORSVulnerableWeb}{475}
\newcommand{\dataOriginsCORSCredsVulnerableWeb}{116}
\newcommand{\dataDomainsCORSCredsVulnerableWeb}{48}
\newcommand{\dataOriginsCORSVulnerableReldom}{267}
\newcommand{\dataDomainsCORSVulnerableReldom}{53}
\newcommand{\dataOriginsCORSCredsVulnerableReldom}{63}
\newcommand{\dataDomainsCORSCredsVulnerableReldom}{27}

\newcommand{\dataOriginsRelaxation}{97}
\newcommand{\dataDomainsRelaxation}{61}
\newcommand{\dataOriginsRelaxationVulnerableReldom}{57}
\newcommand{\dataDomainsRelaxationVulnerableReldom}{29}

\newcommand{\dataCountCookies}{24,924}
\newcommand{\dataOriginsCookies}{15,179}
\newcommand{\dataDomainsCookies}{846}
\newcommand{\dataCountSessionCookies}{8,558}
\newcommand{\dataOriginsSessionCookies}{8,757}
\newcommand{\dataDomainsSessionCookies}{801}
\newcommand{\dataCountCookiesConfidentiality}{3,390}
\newcommand{\dataOriginsCookiesConfidentiality}{5,051}
\newcommand{\dataDomainsCookiesConfidentiality}{687}
\newcommand{\dataCountCookiesIntegrity}{24,689}
\newcommand{\dataOriginsCookiesIntegrity}{14,964}
\newcommand{\dataDomainsCookiesIntegrity}{834}

\newcommand{\dataOriginsPostMessage}{14,045}
\newcommand{\dataDomainsPostMessage}{823}
\newcommand{\dataOriginsPostMessageVulnerableWeb}{14}
\newcommand{\dataDomainsPostMessageVulnerableWeb}{11}
\newcommand{\dataOriginsPostMessageVulnerableReldom}{0}
\newcommand{\dataDomainsPostMessageVulnerableReldom}{0}

\algnewcommand{\LineComment}[1]{\State \(\triangleright\) #1}
\algrenewcommand\algorithmicindent{1.0em}

\newcolumntype{Y}{>{\centering\arraybackslash}X}
\newcolumntype{C}[1]{>{\centering}m{#1}}
\newcolumntype{R}{>{\raggedleft\arraybackslash}X}

\begin{document}

\title{\Large \bf Can I Take Your Subdomain?\\Exploring Related-Domain Attacks in the Modern Web}

\author{
{\rm Marco Squarcina$^1$, Mauro Tempesta$^1$, Lorenzo Veronese$^1$, Stefano Calzavara$^2$, Matteo Maffei$^1$}\\
$^1$TU Wien, $^2$Universit\`a Ca' Foscari Venezia
}

\maketitle

\begin{abstract}
Related-domain attackers control a sibling domain of their target web application, e.g., as the result of a subdomain takeover. Despite their additional power over traditional web attackers, related-domain attackers received only limited attention by the research community. In this paper we define and quantify for the first time the threats that related-domain attackers pose to web application security. In particular, we first clarify the capabilities that related-domain attackers can acquire through different attack vectors, showing that different instances of the related-domain attacker concept are worth attention. We then study how these capabilities can be abused to compromise web application security by focusing on different angles, including: cookies, CSP, CORS, postMessage and domain relaxation. By building on this framework, we report on a large-scale security measurement on the top 50k domains from the Tranco list that led to the discovery of vulnerabilities in \dataVDomains{} sites, where we quantified the threats posed by related-domain attackers to popular web applications.
\end{abstract}

\paragraph*{Note.} This is a preprint of a paper submitted to USENIX Security '21 on 16 Oct 2020.

\section{Introduction}
The Web is the most complex distributed system in the world. Web security practitioners are well aware of this complexity, which is reflected in the threat modeling phase of most web security analyses. When reasoning about web security, one has to consider multiple angles. The \emph{web attacker} is the baseline attacker model that everyone is normally concerned about. A web attacker operates a malicious website and mounts attacks by means of standard HTML and JavaScript, hence any site operator in the world might act as a web attacker against any other service. High-profile sites are normally concerned about \emph{network attackers} who have full control of the unencrypted HTTP traffic, e.g., because they operate a malicious access point. Both web attackers and network attackers are well known to web security experts, yet they do not capture the full spectrum of possible threats to web application security.

In this paper, we are concerned about a less known attacker, referred to as \emph{related-domain attacker}~\cite{Barth08}. A related-domain attacker is traditionally defined as a web attacker with an extra twist, i.e., its malicious website is hosted on a sibling domain of the target web application. For instance, when reasoning about the security of \domain{www.example.com}, one might assume that a related-domain attacker controls \domain{evil.example.com}. The privileged position of a related-domain attacker   endows it  for instance with the ability of compromising cookie confidentiality and integrity, because cookies can be shared between domains with a common ancestor, reflecting the assumption underlying the original web design that related domains are under the control of the same entity. Since client authentication on the Web is mostly implemented on top of cookies, this represents a major security threat.

Despite their practical relevance, related-domain attackers received much less attention than web attackers and network attackers in the web security literature. We believe there are two plausible reasons for this. First, related-domain attackers might sound very specific to cookie security, i.e., for many security analyses they are no more powerful than traditional web attackers, hence can be safely ignored. Moreover, related-domain attackers might appear far-fetched, because one might think that the owner of \domain{example.com} would never grant control of \domain{evil.example.com} to untrusted parties.

Our research starts from the observation that both previous arguments have become questionable and this is the right time to take a second look at the threats posed by related-domain attackers, which are both relevant and realistic. A key observation to make is that a related-domain attacker shares the same \emph{site} of the target web application, i.e., sits on the same registerable domain. The notion of site has become more and more prominent for web security over the years, going well beyond cookie confidentiality and integrity issues. For example, the Site Isolation mechanism of Chromium ensures that pages from different sites are always put into different processes, so as to offer better security guarantees even in presence of bugs in the browser~\cite{ReisMO19}. Moreover, major browsers are now changing their behavior so that cookies are only attached to same-site requests by default, which further differentiates related-domain attackers from web attackers. In the rest of the paper we discuss other (normally overlooked) examples where the privileged position of related-domain attackers may constitute a significant security threat. Finally, many recent research papers showed that \emph{subdomain takeover} is a serious and widespread security risk~\cite{LiuHW16,BorgolteFHKV18}. Large organizations owning a huge number of subdomains might suffer from incorrect configurations, which allow an attacker to make subdomains resolve to a malicious host. This problem also received attention from the general media~\cite{Zdnet} and the industry~\cite{Biasini15}. Though these studies proved that related-domain attackers are a realistic threat, they never quantified their impact on web application security at scale.

\subsection*{Contributions}
In the present paper, we perform the first scientific analysis of the dangers represented by related-domain attackers to web application security. In particular:
\begin{enumerate}
    \item We introduce a fine-grained definition of \rda{}  that captures the capabilities granted to such attacker according to the position they operate and the associated web security threats. In particular, we systematize  the attack vectors that an attacker can exploit to gain control of a domain, and we present the attacks that can be launched from that privileged position, discussing the additional gain with respect to a traditional web attacker (§\ref{sec:threat}).
    
	\item We implement a toolchain to evaluate the dangers that related-domain attackers can pose to web application security. Our toolchain builds on top of an analysis module for subdomain takeover, which significantly improves over previous results~\cite{LiuHW16}. We use the output of this module to perform automated web application security analyses along different angles, including: cookies, CSP, CORS, postMessage and domain relaxation (§\ref{sec:methodology}).
    
    \item We report on experimental results established through our toolchain. In particular, we enumerate 26M subdomains of the top 50k registrable domains and discover practically exploitable vulnerabilities in \dataVDomains{} domains, including major websites like \domain{cnn.com}, \domain{nih.gov}, \domain{harvard.edu}, and \domain{cisco.com}. We also study the security implications of \dataTotServices{} third-party service providers and dynamic DNS, and present a novel subdomain hijacking technique that resulted in a bug bounty of \$1,000. Importantly, we quantify for the first time the impact of these vulnerabilities on web application security, concluding that related-domain attackers have an additional gain compared to web attackers that goes beyond well studied issues on cookies (§\ref{sec:results}).
\end{enumerate}

We have responsibly disclosed the identified vulnerabilities to the respective site operators. For space reasons, the results of the notification process are shown in Appendix~\ref{appendix:disclosure}.

\section{Background}
\paragraph{DNS Resolution.}
DNS is a protocol that stands at the core of the Internet~\cite{RFC1035}. It translates mnemonic domain names to IP addresses used by the underlying network layer to identify the associated resources. The translation process, called \emph{DNS resolution}, is done transparently to applications. For instance, when a browser attempts to visit a fully qualified domain name (FQDN), such as \domain{www.example.com}, a local resolver tries to find the corresponding DNS record in a local cache. If no match is found, the local resolver forwards the request to one of the DNS servers designated by the operating system. In case the DNS server has no information on the requested domain name, it initiates the recursive resolution from the root DNS server until the \emph{authoritative} DNS server for the domain is reached, following the \emph{subdomain} hierarchy of DNS system. Eventually, the authoritative DNS server returns to the client a set of Resource Records (RRs) with the following format: \emph{name}, \emph{TTL}, \emph{class}, \emph{type}, \emph{data}.

\begin{figure}[t]
\begin{Verbatim}[fontsize=\footnotesize, frame=single]
;; ANSWER SECTION:  
www.shop.fox.com.     300 IN CNAME shops.myshopify.com.
shops.myshopify.com. 3600 IN A     23.227.38.64
\end{Verbatim}
\vspace{-10pt}
\caption{Example of a DNS resolution.}\label{fig:dns-answer}
\end{figure}

Figure~\ref{fig:dns-answer} provides an example of a DNS resolution for \domain{www.shop.fox.com}, as shown by the \texttt{dig} utility. The first line tells that \domain{www.shop.fox.com} is a \texttt{CNAME} record type, pointing to the \emph{canonical name} \domain{shops.myshopify.com}. The canonical name is then resolved into an \texttt{A} record with the data field containing the IP. The result of the resolution process is the translation of the name \domain{www.shop.fox.com} into the IP \texttt{23.227.38.64}. A list of relevant DNS record types is summarized in Table~\ref{tab:dns-types}.

\begin{table}[t]
    \footnotesize
	\caption{Main DNS record types.}
    \label{tab:dns-types}
    \renewcommand\tabularxcolumn[1]{m{#1}}
	\begin{tabularx}{\linewidth}{@{}lX@{}}
	\toprule
        Record Type & Description \\
        \midrule
        \texttt{A} & Returns the IPv4 address of a domain\\
        \texttt{AAAA} & Returns the IPv6 address of a domain\\
        \texttt{CNAME} & Maps an alias name to the canonical domain name\\
        \texttt{NS} & Defines the authoritative DNS record for a domain\\
        \texttt{CAA} & Specifies the allowed certificate authorities for a domain\\
        \bottomrule
	\end{tabularx}
\end{table}

DNS also supports \emph{wildcard} RRs with the label \texttt{*}, such as \domain{*.example.com}. Wildcard RRs are not matched if an explicit RR is defined for the requested domain name. In general, wildcard RRs have a lower priority than standard RRs~\cite{RFC4592}. For instance, given a wildcard \texttt{A} record \domain{*.example.com} and an \texttt{A} record for \domain{foo.example.com}, requests to \domain{bar.example.com} and \domain{baz.bar.example.com} will match the wildcard, as opposed to \domain{foo.example.com} and \domain{baz.foo.example.com}.

\paragraph{Public Suffix List.}
While DNS defines the hierarchical structure of domain names, the Public Suffix List (PSL) is a catalog of domain suffixes controlled by registrars~\cite{PSL}. In contrast to Top-Level Domains (TLDs) that are defined in the Root Zone Database~\cite{RZD}, such as \texttt{.com}, \texttt{.org}, \texttt{.net}, the suffixes listed in the PSL are called \emph{effective TLDs} (eTLDs) and they define the boundary between names that can be registered by individuals and private names. A domain name having just one label at the left of a public suffix is commonly referred to as \emph{registrable domain}, \emph{eTLD+1} or \emph{apex domain}. Domains sharing the same eTLD are often referred to as belonging to the same \emph{site}.

Cookies are scoped based on the definition of site, i.e., subdomains of the same site can share cookies (\emph{domain cookies}) by setting their \texttt{Domain} attribute to a common ancestor. This attribute can never be set to a member of the PSL: for instance, since \domain{github.io} is in the PSL, \domain{foo.github.io} is not allowed to set cookies for \domain{github.io}. This means that there is no way to share cookies between different GitHub Pages hosted sites.

\section{The Related-Domain Attacker}
\label{sec:threat}
\begin{figure}[t!]
    \centering
    \includegraphics[width=0.45\textwidth]{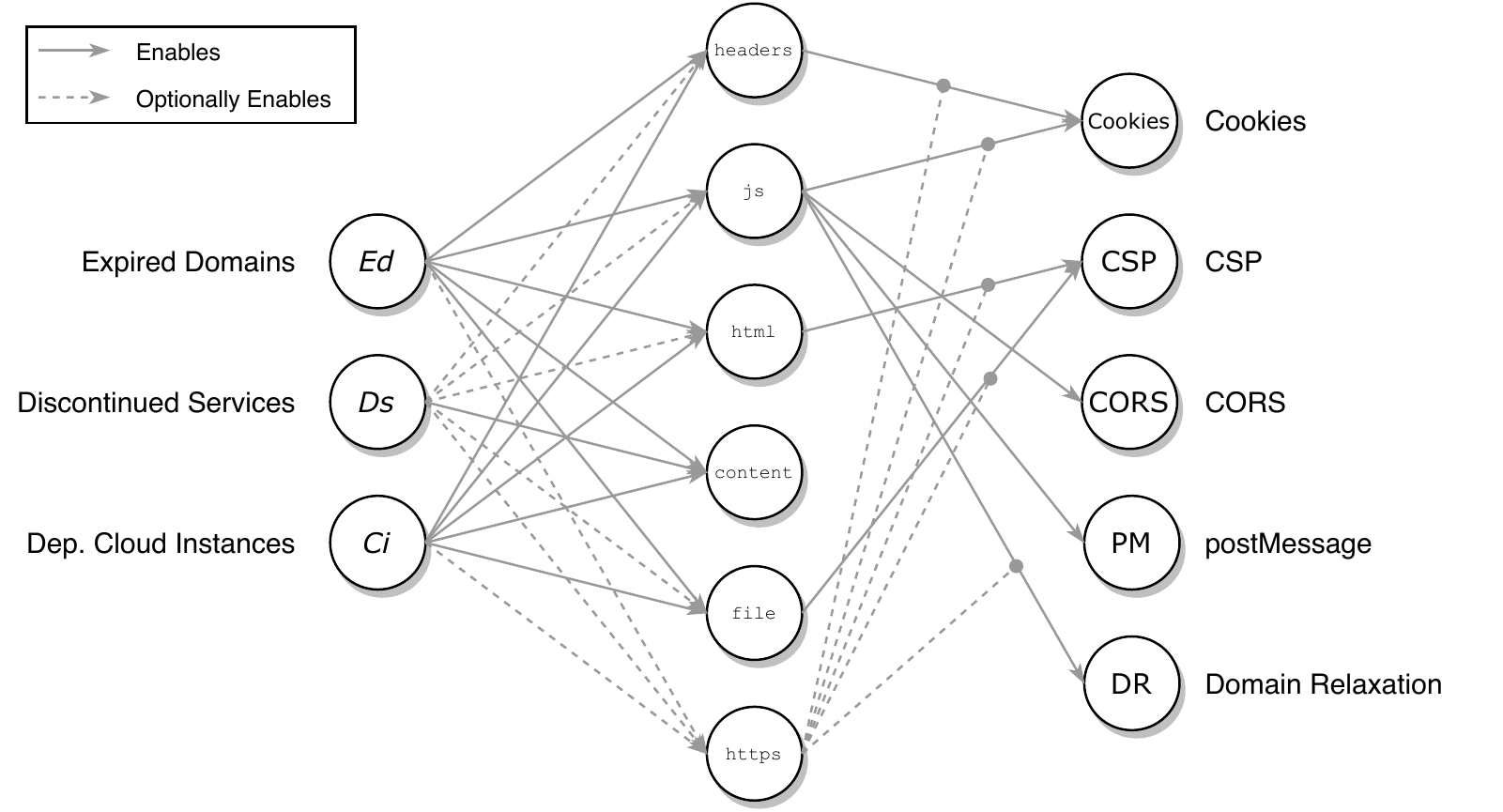}
    \caption{Summary of related-domain attacker instances.}
    \label{fig:summarythreat}
\end{figure}

We revise the threat model of the \rda{} at the light of the direction that the Web has taken in the recent years. In particular, we systematize for the first time the different attack vectors that can be exploited to escalate to a related-domain position, the acquired capabilities, and the associated web security threats, as summarized in Figure \ref{fig:summarythreat} and presented below. This systematization allows for a quantification of the  \rda{} problem, which we conduct in §\ref{sec:results} by a large-scale measurement in the wild.

\subsection{Threat Model}\label{subsec:threat-model}
In its original definition, the \rda{} operates a malicious website that is hosted on a \emph{related-domain} of the target website~\cite{Bortz11}. Two domains are \emph{related} if they share a suffix that is not included in the PSL. For instance, consider the target site \domain{example.com}: all its subdomains are related to the target, as well as being related to each other.

There are several ways for attackers to position themselves on a related-domain of the target. DNS misconfigurations are well known to cause subdomain takeovers~\cite{LiuHW16,BorgolteFHKV18} with consequences ranging from altering the content of a page to full host control. Organizations frequently assign a subdomain of their corporate domain to their users, who could maliciously take advantage of this implicit trust. Additionally, vulnerable web applications can be infiltrated to increase the privileges of attackers interested in exploiting their related domains.

As we elaborate in the following, the attack vector exploited to acquire a related-domain position is not a detail, but has an impact on the \emph{capabilities} granted to the attacker. While full control of the host grants the attacker with the ability to configure the web server to host arbitrary content, other attack scenarios only grant more limited power. For example, exploiting a reflected XSS on a subdomain of a company poses several restrictions on the actions that can be undertaken by the attacker. This motivates the need for a new, fine-grained definition of \rda{}, which  precisely characterizes its power based on the acquired capabilities. In  §\ref{subsec:abusing-related-domains}, we map concrete attack vectors to the set of capabilities (see Table~\ref{tab:capabilities}) that the attacker may acquire when escalating to a related-domain position. In §\ref{subsec:attacks}, we link  such capabilities to web security threats. This gives rise to a granular framework defining different instances of the related-domain attacker.

\begin{table}[t]
        \footnotesize
	\caption{Capabilities of the \rda{}.}
	\label{tab:capabilities}
	\begin{tabularx}{\linewidth}{@{}lX@{}}
	\toprule
        Capability & Description \\
        \midrule
	\capab{headers} & access and modify HTTP headers\\
	\capab{js} & arbitrary Javascript code execution\\
        \capab{html} & alter the markup of the website with the exclusion of javascript \\
	\capab{content} & alter the textual content of the website with the exclusion of embed tags, frames and javascript code \\
	\capab{file} & host arbitrary files\\
        \capab{https} & operate a website under HTTPS with a valid certificate\\
        \bottomrule
	\end{tabularx}
        \ \\ \ {\footnotesize \textbf{Note:} \capab{js} subsumes both \capab{html} and \capab{content}, since it is possible to edit the DOM by using JavaScript. Similarly, \capab{html} subsumes \capab{content}. }
\end{table}

\subsection{Abusing Related Domains}\label{subsec:abusing-related-domains}
We made an effort to provide a comprehensive characterization of the attack vectors that can be exploited to acquire a related-domain position along with the associated capabilities. While some of these attack vectors have been already analyzed in the literature in isolation (e.g., dangling DNS records~\cite{LiuHW16} and domain shadowing~\cite{Biasini15,LiuLDWLD17}), it is the first time they are systematized to cover the possible abuses which enable escalation to a related-domain position.

\subsubsection{Dangling DNS Records}\label{subsubsec:dangling-dns-records}
Dangling DNS records refer to records in the authoritative DNS servers of a domain that point to expired resources. Dangling DNS records should be purged right away after releasing the pointed resource. Unfortunately, this practice is often overlooked resulting in dangling DNS records to persist indefinitely. Possible reasons include lack of communication between the person who releases the resource and the domain owner or when the pointed resource expires automatically after a certain period of time, passing unnoticed. A dangling DNS record is considered vulnerable if a third-party other than the intended one can control the expired resource~\cite{LiuHW16}.

\paragraph{Expired Domains.} A DNS CNAME record maps a domain name (alias) to another one called canonical name. If the canonical name is expired, a third party can simply register the domain and serve arbitrary content under the alias domain. Attackers exploiting this vulnerability have full control of the host and generally can rely on all the capabilities listed in our framework. One exception is \capab{https} in presence of a CAA DNS record~\cite{RFC8659}: this  record defines a list of Certificate Authorities (CAs) which are allowed to issue certificates for a given domain, possibly preventing attackers to rely on automated CAs like Let's Encrypt~\cite{AasEtAl19}.

\paragraph{Discontinued Services.} Third-party services are widely used to extend the functionalities of a website. Although a site or page hosted by the service provider is automatically instantiated after registration, domain owners can integrate the service by associating a custom domain to make the platform reachable under a subdomain of the organization, e.g., \domain{blog.example.com} for a blogging service like Wordpress and \domain{shop.example.com} in the case of an e-shop like Shopify. Setting up a custom domain typically requires (i) certain DNS records such as \texttt{A/AAAA}, \texttt{CNAME} or \texttt{NS}, to point to a resource controlled by the service provider and (ii) to claim the domain ownership in the service configuration interface. If the service provider does not verify the domain ownership, DNS records pointing to discontinued services can be abused by any illegitimate third-party that registers an account on the service provider and binds the domain to the attacker's account~\cite{LiuHW16}.

In addition, we observe that dangling records can also occur due to the presence of DNS wildcard. Consider, for example, a site operator configuring a DNS wildcard such as \domain{*.example.com} pointing to a service provider IP to enable multiple websites to be hosted under subdomains of \domain{example.com}. An attacker could bind a subdomain of their choice, e.g., \domain{evil.example.com}, to a new account on the service provider. Surprisingly, we discovered that some service providers do not verify the ownership of a subdomain, even if the parent domain has been already mapped to an existing account. In practice, this allows an attacker to claim \domain{evil.project.example.com} also in presence of a legitimate binding for \domain{project.example.com}. Even worse, we found that some service providers perform an automatic redirection from the \texttt{www} prefix to the associated domain, without preventing the \texttt{www} subdomain from being mapped to a different account.
Attackers' capabilities vary depending on the platform, and range from altering the content of a single page to full host control. We refer to §\ref{sec:results} for the result of a throughout security investigation conducted on \dataTotServices{} service providers.

\paragraph{Deprovisioned Cloud Instances.} The ephemeral nature of resources allocated in Infrastructure as a Service (IaaS) environments is known to facilitate the spread of dangling DNS records. DNS records pointing to available IP addresses in the cloud can be abused by a determined attacker who rapidly allocates IP addresses in order to control the target of the dangling DNS record~\cite{LiuHW16,BorgolteFHKV18}. Similarly to expired domains, the presence of a CAA DNS record in a parent domain could hinder the capability of obtaining a valid TLS certificate.
  
\subsubsection{Corporate Networks and Roaming Services}\label{subsubsec:corporate-networks}
Large organizations often assign fully qualified domain names (FQDNs) to hosts  and devices in their network. This practice allows system administrators to statically reference resources in the network, irrespectively of the assignment of IP addresses that may change over time. Although hosts might be inaccessible from outside of the organization network due to an internal resolver or firewall restrictions,  internal users are put in a \rda{} position.

Local institutions providing roaming services are similarly prone to the same issue, with the difference that users being assigned a FQDN might not be related to the service provider. This is the case of eduroam, a popular international education roaming service that enables students and researchers to have a network connection provided by any of the participating institutions. As a novel insight, we discovered that system integrators at some local institutions are assigning eduroam users a subdomain of the main institution, such as \domain{ip1-2-3-4.eduroam.example.com}, where \domain{1.2.3.4} is a placeholder for the public IP assigned to the user connected to the eduroam network. This practice ultimately promotes any eduroam user to a \rda{} with full control of the host that is pointed by the DNS record. Firewall restrictions might hinder complete visibility on the Internet of the personal device of the user. Still, users' devices might be accessible within the institution network, as in our university.

\subsubsection{Hosting Providers and Dynamic DNS Services}
Many service providers allow users to create websites under a specific subdomain. For instance, GitHub Pages enables the creation of websites at \domain{<username>.github.io}, where username is the account or the organization name on GitHub that is instantiating the website. Subdomains hosting user-supplied content are not related to each other if the parent domain is included in the PSL, as in the case of GitHub Pages. Unfortunately, several service providers that we reviewed, did not include their domains in the PSL, turning any of their users into a \rda{} for all the websites hosted on the same platform.

A similar consideration applies to dynamic DNS providers. The race to offer a huge variety of domains under which users can create their custom subdomains, made it unfeasible for certain providers to maintain a list of entries in the PSL. The FreeDNS service~\cite{FreeDNS} pictures well the problem, with 52,443 domains offered and a declared user base of 3,448,806 active users as of October 2020, who are in a \rda{} position to all the subdomain and domains of the network, since none of them has been added to the PSL.

While in the case of hosting and service providers, the capabilities granted to the attacker largely depends on the specific service (see §\ref{subsubsec:analysis-services} for more details), a dynamic DNS service allows users to point a DNS record to a host they fully control, capturing all the capabilities discussed in Table~\ref{tab:capabilities}.

\subsubsection{Compromised Hosts/Websites}\label{subsec:compromised-sites}
Aside from scenarios in which attackers gain control of a resource that is either abandoned or explicitly assigned to them, another way to obtain a \rda{} position is the exploitation of vulnerable hosts and websites. Intuitively, attackers achieving code execution on the vulnerable application have capabilities ranging from serving arbitrary content to full host control. If the exploited vulnerability is an XSS, attackers could take advantage of the ability to execute JavaScript code from a privileged position to escalate the attack against a more sensitive website.

Furthermore, attackers have been found employing a technique called \emph{domain shadowing}~\cite{Biasini15,LiuLDWLD17} to illicitly access the DNS control panel of active domains to distribute malware from arbitrary subdomains. Alowaisheq~\etal{} recently discovered that stale \texttt{NS} records~\cite{Alowaisheq20} could be also abused by attackers to take control of the DNS zone of a domain to create arbitrary DNS records. Controlling the DNS of a domain is the highest privileged setting for a \rda{}s, since they can point subdomains to hosts they fully control and reliably obtain TLS certificates.

\subsection{Web Threats}\label{subsec:attacks}
We identify for the first time a comprehensive list of web security threats posed by \rda s, discussing in particular the scenarios where a \rda{}  might have an advantage over traditional web attackers. While threats to cookie confidentiality and integrity from \rda s have been thoroughly explored in the literature~\cite{ZhengJLDCWW15}, many others have not been explored so far.

\subsubsection{Inherent Threats}
Related-domain attackers sit on the same site of their target web application. This is weaker than sharing the same \emph{origin} of the target, which is the traditional web security boundary, yet it suffices to abuse the trust put by browser vendors and end users on same-site content. We discuss examples below.

\paragraph{Trust of End Users.} End users might trust subdomains of sites they are familiar with more than arbitrary external sites. For example, they might be more willing to insert their Facebook password on \domain{abc.facebook.com} rather than on \domain{facebooksite.com}. This is particularly dangerous on some mobile browsers, which display only the rightmost part of the domain due to the smaller display size, hence a long subdomain might erroneously look like the main site. Subdomains can similarly abuse trust to distribute malware or other types of dangerous content~\cite{LiuLDWLD17}.

\paragraph{Site Isolation.} Site Isolation is a browser architecture first proposed and implemented by the Google Chrome browser, which treats different sites as separate security principals requiring dedicate rendering processes~\cite{ReisMO19}. Hence, these processes can access sensitive data for a single site only, which mitigates the leakage of cross-origin data via memory disclosure and renderer exploits. As acknowledged in the original Site Isolation paper~\cite{ReisMO19}, ``cross-origin attacks within a site are not mitigated'', hence related-domain attackers can void the benefits of this security architecture.

\paragraph{Same Site Request Forgery.}
The introduction of \emph{same-site cookies}~\cite{RFC6265bit06} and the recent enforcement of this security feature by default on major browsers~\cite{ChromiumSameSite, MozillaSameSite} received high praise, as an effective countermeasure against CSRF~\cite{CSRFDead}. In the absence of other countermeasures~\cite{Barth08}, the restrictions introduced by same-site cookies are voided by a \rda{}, who can mount a \emph{same-site} request forgery attack just by including an HTML element pointing to the target website in one of their web pages.

\subsubsection{Cookie Confidentiality and Integrity}\label{subsubsec:attacks-cookies}
Cookies can be issued with the \texttt{Domain} attribute set to an ancestor of the domain setting them, so as to share them with all its subdomains. For example, \domain{good.foo.com} can issue a cookie with the \texttt{Domain} attribute set to \domain{foo.com}, which is sent to both \domain{good.foo.com} and \domain{evil.foo.com}. Hence, \rda{}s can trivially break \emph{cookie confidentiality} and abuse of stolen cookies~\cite{ZhengJLDCWW15}, e.g., to perform session hijacking. The \texttt{Domain} attribute poses risks to \emph{cookie integrity} too:  \domain{evil.foo.com} can set cookies for \domain{good.foo.com}, which can be abused to mount attacks like \emph{session fixation}~\cite{Kolsek02}. Note that the integrity of host-only cookies is at harm too, because a \rda{} can mount \emph{cookie shadowing}, i.e., set a domain cookie with the same name of a host-only cookie to confuse the web server~\cite{ZhengJLDCWW15}.

Site operators can defend against such threats by careful cookie management. For example, they can implement (part of) the session management logic on top of host-only cookies, which are not disclosed to related-domain attackers. Moreover, they can use the \texttt{\_\_Host-} prefix to ensure that security-sensitive cookies are set as host-only, thus ensuring their integrity against related-domain attackers.

\paragraph{Capabilities.} The capabilities required by a \rda{} to break the confidentiality of a domain cookie depend on the flags enabled for it: if the cookie is \texttt{HttpOnly}, it cannot be exfiltrated via JavaScript and the \texttt{headers} capability is needed to sniff it; otherwise, just one between \capab{headers} and \capab{js} suffices. Note that, if the \texttt{Secure} flag is enabled, the cookie is only disclosed to pages loaded over HTTPS, hence the \capab{https} capability is also required. As to integrity, all cookies lacking the \texttt{\_\_Host-} prefix have low integrity against a \rda{} with the \capab{headers} or \capab{js} (via \emph{cookie tossing}~\cite{GithubCookies}) capabilities. There is just one exception: cookies using the \texttt{\_\_Secure-} prefix have low integrity only against \rda{} which additionally have the \capab{https} capability, since these cookies can only be set over HTTPS.

\subsubsection{Bypassing CSP}\label{subsubsec:attacks-csp}
Content Security Policy (CSP) is a client-side defense mechanism originally designed to mitigate the dangers of content injection and later extended to account for different threats, e.g., click-jacking. CSP implements a white-listing approach to web application security, whereby the browser behavior on CSP-protected web pages is restrained by binding \emph{directives} to sets of \emph{source expressions}, i.e., a sort of regular expressions designed to express sets of origins in a compact way. To exemplify, consider the following CSP:
\begin{Verbatim}[fontsize=\small,frame=single]
script-src foo.com *.bar.com;
frame-ancestors *.bar.com;
default-src https:
\end{Verbatim}
This policy contains three directives, \texttt{script-src}, \texttt{frame-ancestors} and \texttt{default-src}, each bound to a set of source expressions like \texttt{foo.com} and \texttt{*.bar.com}. It allows the protected page to:
\begin{enumerate*}[label={(\roman*)}]
	\item include scripts from \texttt{foo.com} and any subdomain of \texttt{bar.com};
	\item be included in frames opened on pages hosted on any subdomain of \texttt{bar.com};
	\item include any content other than scripts over HTTPS connections with any host.
\end{enumerate*}
Moreover, the policy implicitly forbids the execution of inline scripts and event handlers, as well as the invocation of a few dangerous functions like \texttt{eval}, which strongly mitigates XSS.

Since the syntax of source expressions naturally supports the whitelisting of any subdomain of a given parent, related-domain attackers represent a major threat against the security of CSP. For example, if an attacker could get control of \texttt{vuln.bar.com}, then she would be able to bypass most of the protection put in place by the CSP above. In particular, the attacker would be able to exploit a content injection vulnerability on the CSP-protected page to load and execute arbitrary scripts from \texttt{vuln.bar.com}, thus voiding XSS mitigation. Moreover, the attacker could frame the CSP-protected page on \texttt{vuln.bar.com} to perform click-jacking attacks. To avoid these threats, site operators should carefully vet the subdomains included in their CSP whitelists.

\paragraph{Capabilities.}
A \rda{} requires the capability to upload arbitrary files on the website under its control to void the protection offered by CSP against content inclusion vulnerabilities, with the only notable exception of frame inclusion which requires only the \capab{html} capability. For active contents~\cite{MixedContent}, i.e., those that may have access to the DOM of the page, the attacker also needs the \capab{https} capability if the target page is hosted over HTTPS.
Regarding click-jacking protection, the attacker just requires the \capab{html} capability in order to include the target website on a page under her control.

\subsubsection{Abusing CORS}
Cross-Origin Resource Sharing (CORS) is the standard approach to relax the restrictions enforced by SOP on accesses to cross-origin data. To exemplify the importance of CORS, consider a medical service at \texttt{https://www.foo.com}, which needs to fetch electronic health records from the subdomain \texttt{patients.foo.com} using an AJAX request. Since \texttt{https://www.foo.com} and \texttt{https://patients.foo.com} are two different origins, SOP prevents the medical service from reading the content of the AJAX response to avoid potential privacy leaks.

Luckily, CORS allows site operators to relax this security restriction: in particular, \texttt{https://patients.foo.com} can inspect the \texttt{Origin} header of the AJAX request to learn that the request comes from \texttt{https://www.foo.com}. Since this is an authorized party, \texttt{https://patients.foo.com} can set the \texttt{Access-Control-Allow-Origin} header to the value \texttt{https://www.foo.com}, so as to instruct the browser into granting cross-origin access to the AJAX response.
As an additional layer of protection, the server must set also the \texttt{Access-Control-Allow-Credentials} to \texttt{true} if the cross-site request includes credentials, e.g., cookies, since these requests are more likely to include security-relevant contents.

Related-domain attackers can abuse CORS to bypass the security restrictions put in place by SOP when the aforementioned server-side authorization checks are too relaxed, i.e., read access is granted to arbitrary subdomains.
For example, if \texttt{https://patients.foo.com} was willing to grant cross-origin access to any subdomain of \domain{foo.com} besides \domain{www.foo.com}, a related-domain attacker could get unconstrained access to electronic health records.
To avoid these threats, site operators should be careful in the security policy implemented upon inspection of the \texttt{Origin} header, e.g., restricting access just to few highly trusted subdomains.

\paragraph{Capabilities.}
To exploit CORS misconfigurations, an attacker needs the \capab{js} capability to issue requests via JavaScript APIs like \texttt{fetch} and access the contents of the response.

\subsubsection{Abusing postMessage}
The postMessage API supports cross-origin communication across windows in modern browsers (e.g., between frames or between a page and the popup opened by it). The sender can invoke the \texttt{postMessage} method of the target window to transmit a message, possibly restricting the origin of the receiver. The receiver, in turn, can use event handlers to listen for the \texttt{message} event and process incoming messages.

Despite its apparent simplicity, the postMessage API should be used with care, as shown by prior research~\cite{SonS13,SteffensS20}. In particular, when sending confidential data, one should always specify the origin of the intended receiver in the \texttt{postMessage} invocation. When receiving data, instead, one should check the origin of the sender (via the \texttt{origin} property of the received message) and appropriately sanitize the content of the message before processing it.

Related-domain attackers can undermine web application security when site operators put additional trust in subdomains.
In particular, related-domain attackers can try to abuse of their position to void the aforementioned origin checks and communicate with inattentive receivers who process messages in an unsafe way, e.g., they are provided as input to \texttt{eval} or stored in a cookie, opening the way to session hijacking attacks. Site operators can defend against such attacks by avoiding overprivilege, i.e., by carefully vetting authorized subdomains for frame communication.

\paragraph{Capabilities.} An attacker requires scripting capabilities (\capab{js}) to open a new tab containing the vulnerable page and communicate with it via the postMessage API.

\subsubsection{Abusing Domain Relaxation}
\label{subsec:domain-relaxation}
Domain relaxation is the legacy way to implement communication between windows whose domains share a common ancestor. Assume that a page at \domain{a.foo.com} opens a page at \domain{b.foo.com} inside a frame. Besides using the postMessage API as explained, the two frames can communicate by relaxing their \domain{document.domain} property to a common ancestor. In this case, both frames can set such property to \domain{foo.com}, thus moving into a same-origin position.\footnote{We assume here that the two frames share the same protocol and port.} After that, SOP does not enforce any isolation between the two frames, which can communicate by writing on each other's DOM or in the web storage. Note that \domain{foo.com} must explicitly set the \texttt{document.domain} property to \domain{foo.com} if it is willing to engage in the domain relaxation mechanism, although this is apparently a no-op.

Domain relaxation can be abused by related-domain attackers, who can look for pages which are willing to engage in such dangerous communication mechanism and abuse it. In particular, when the attacker moves into a same-origin position, SOP does not provide any protection anymore, which voids any confidentiality and integrity guarantee.
Websites that are willing to communicate with a selected list of related-domain should refrain from using this mechanism, which is deemed as insecure, and should implement cross-origin communication on top of the \texttt{postMessage} API.

\paragraph{Capabilities.} 
Besides the \capab{js} capability needed to perform the relaxation and access the DOM of the target page, the attacker needs to setup her attack page on the same protocol of the target, hence the \capab{https} capability may also be required.

\section{Analysis Methodology}
\label{sec:methodology}
\begin{figure}
	\center
	\includegraphics[width=\columnwidth]{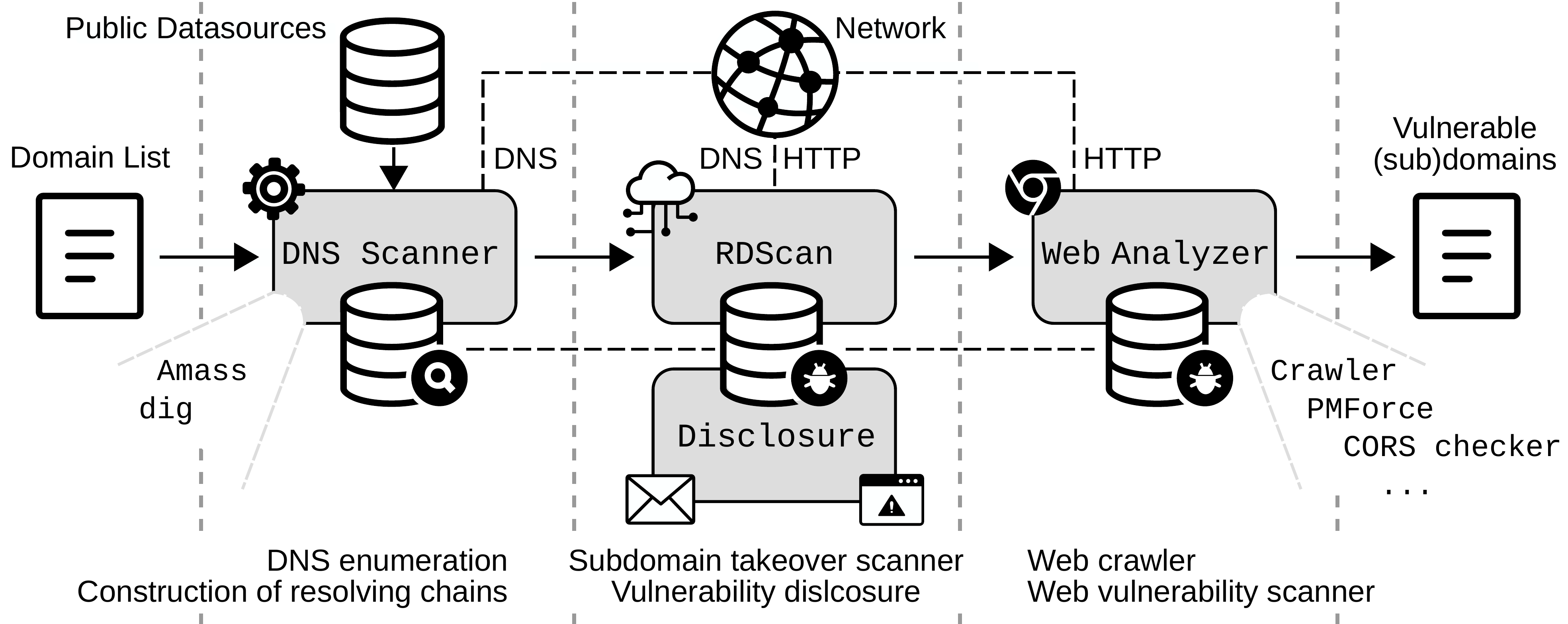}
	\caption{Vulnerability scanning pipeline.}\label{fig:pipeline}
\end{figure}

We performed a large-scale vulnerability assessment to measure the pervasiveness of the threats reported in this work, first by identifying subdomains of prominent websites that can be abused by a \rda{} exploiting dangling DNS records, and second by evaluating the security implications on web applications hosted on related domains of the vulnerable websites. Our methodology is based on the pipeline summarized in Figure~\ref{fig:pipeline} and further described in this section. We conducted the analysis driven by the principle of minimizing the impact on the analyzed websites. This directly translated into lowering the network traffic by relying on public data sources as much as possible, having a positive repercussion on the cost required to perform the experiments.

\subsection{DNS Data Collection and RDScan}\label{subsec:dns-data-collection}
We enumerated the subdomains of the top 50k domains in the Tranco list~\cite{LePochat19} from March 2020.\footnote{https://tranco-list.eu/list/ZKYG/1000000}. The enumeration phase was based on \texttt{amass}~\cite{Amass}, a state of the art information gathering tool backed by the OWASP project. The tool supports several techniques to maximize the chances of discovering subdomains of a target. In our configuration, we extracted subdomains using the following approaches:
\begin{enumerate*}[label=(\roman*)]
	\item fetch data from publicly available sources, such as Censys~\cite{Censys}, certificate transparency logs~\cite{Crtsh}, search engines, etc.;
	\item attempt DNS zone transfer~\cite{RFC5936} to obtain the complete list of RRs defined for a certain DNS zone;
	\item inspect fields of TLS certificates, \eg{}, \texttt{Subject Alternative Name} and \texttt{Common Name}~\cite{RFC5280}.
\end{enumerate*}
To speed up the enumeration phase and lower the number of network requests, we avoid bruteforcing DNS resolvers against domain name wordlists. Similarly, we explicitly disabled the resolution of subdomain alterations.

We modified \texttt{amass} to compute the DNS \emph{resolving chains} of all the domains obtained in the previous step. Similarly to~\cite{LiuHW16}, we define a resolving chain as list of DNS RRs in which each element is the target of the previous one, starting from a DNS record of type \texttt{A/AAAA}, \texttt{CNAME} or \texttt{NS}. We do not consider \texttt{MX} records because we focus on web attacks in this study. For \texttt{CNAME} and \texttt{NS} records, we recursively perform a DNS resolution until an \texttt{A/AAAA} RR is detected. Unterminated DNS resolving chains can occur in presence of a record pointing to an unresolvable resource or due to the abrupt termination of \texttt{amass} due to the execution timeout of 5 minutes. To ensure the correctness of the results, we recompute unterminated DNS resolving chains using the \texttt{dig} utility.

Starting from the set of 50k domains in the Tranco list, our framework identified around 26 million valid subdomains. In a previous study, Liu~\etal{}~\cite{LiuHW16} used a relatively small wordlist of 20,000 entries to find possible subdomains of the Alexa top 10k list, 2,700 .edu domains, and 1,700 .gov domains. Compared to their work, our domain selection is penalized given that we do not restrict to specific TLD zones. For instance, .edu domains typically have a high number of subdomains in contrast to other categories (see §\ref{subsubsec:characterization}). Nevertheless, our results outperform the findings of Liu~\etal{} by discovering on average 13 times more subdomains due to the higher coverage of enumeration techniques in our scan.

After populating a database with the DNS records of the discovered subdomains, the framework performs the identification of dangling records and verifies the precondition for a successful subdomain takeover to exclude false positives from our analysis. This component, that we call \emph{RDScan}, has three different modules that test for the presence of the vulnerable scenarios mentioned in §\ref{subsubsec:dangling-dns-records}.

Expired domains are discovered by finding resolving chains that point to a domain that can be purchased at any registrar. For discontinued services, RDScan supports testing procedures for \dataTotServices{} service providers and dynamic DNS services. Pruning false positives required in few cases to temporarily associate a vulnerable subdomain to our testing account. In case the mapping was successful, we released the acquired resource right after being checked. We discuss the ethical implications of our analysis in Appendix~\ref{appendix:disclosure}. Finally, we used an heuristic to detect potentially deprovisioned cloud instances on 6 major cloud providers: Amazon AWS, GoogleCloud Platform, Microsoft Azure, Hetzner Cloud, Linode, and OVHcloud. Our approach is similar to~\cite{LiuHW16,BorgolteFHKV18}, but we relied on a publicly available dataset~\cite{Rapid7} for the \emph{liveness probe} of the cloud resource. A detailed description of RDScan is presented in Appendix~\ref{appendix:rdscan}.

\subsection{Web Analyzer}

Our web security analysis aims at quantifying the number of domains hosting web applications that can be exploited by taking over the vulnerable domains discovered by RDScan.
In particular, for every apex domain with at least one vulnerable subdomain, we selected from the CommonCrawl dataset~\cite{CommonCrawl} the list of 200 most popular related-domains according to the Pagerank score~\cite{BrinP98}.
From the homepage of these domains, we extracted the same-origin links that appear in the HTML code. We picked (up to) 5 of these URLs and the homepage for every related-domain as the target of our web analysis, which we visited using an instance of the Chromium browser instrumented with puppeteer.
In the following we discuss the data collection process and the security analyses we have conducted to identify the security threats discussed in §\ref{subsec:attacks}, while we postpone the discussion of the results to §\ref{sec:results}.

\subsubsection{Analysis of Cookies}

Thanks to puppeteer APIs, we collect the cookies set by the page via HTTP headers or JavaScript and we adopt the heuristics proposed by Bugliesi \emph{et al.}~\cite{BugliesiCFK15} to restrict our analysis to (pre-)session cookies, i.e., cookies that may be relevant for the management of user sessions.

Our goal is to identify cookies affected by confidentiality or integrity issues. In particular, we flag a cookie as affected by confidentiality issues if, among the related domains vulnerable to takeover, there exists a domain $d$ such that:
\begin{itemize}[noitemsep]
	\item $d$ is a subdomain of the \texttt{Domain} attribute of the cookie;
	\item by taking over $d$, the attacker has acquired the capabilities required to leak the cookie.
\end{itemize}
We mark a cookie as affected by integrity issues if:
\begin{itemize}[noitemsep]
	\item the name of the cookie does not start with \texttt{\_\_Host-};
	\item we identified a vulnerable domain that grants the capabilities required to perform cookie forcing.
\end{itemize}
The capabilities required to perform these attacks depend on the security flags assigned to the cookie and the usage of cookie prefixes (see §\ref{subsubsec:attacks-cookies}). For instance, to compromise integrity either the capability \capab{js} or \capab{headers} is required and, if the prefix \texttt{\_\_Secure-} is used, \capab{https} is also necessary.

\subsubsection{Analysis of CSP policies}

For this analysis we implemented an evaluator of CSP policies according to the draft of the latest CSP version~\cite{CSPv3}, which is currently supported by all major browsers.
This is not a straightforward task, due to the rich expressiveness of the policy and various aspects that have been introduced into the specification for compatibility purposes across different CSP versions, e.g., for scripts and styles, the \texttt{'unsafe-inline'} keyword, which whitelists arbitrary inline contents in a page, is discarded when hashes or nonces (which allow only selected inline contents) are also specified.

In our analysis of CSP we focus on the protection offered against click-jacking and the inclusion of active contents~\cite{MixedContent}, i.e., resources that have access to (part of) the DOM of the embedding page and thus could be abused to disrupt the integrity of a web page. This class of contents include scripts, stylesheets, objects and frames.

For each threat considered in our analysis, first we check if the policy shipped with a page is unsafe with respect to any web attacker.
Intuitively, a policy is unsafe with respect to a certain threat if it allows for inclusion of contents from any host (or framing by any host, when focusing on click-jacking protection). For scripts and styles, the policy is also deemed as unsafe if arbitrary inline contents are whitelisted.

If the policy is considered safe, we check whether a related-domain of the page under analysis (excluding the registrable domain) is whitelisted by all the policies and, in such a case, we say that the CSP is potentially exploitable by a related-domain. 
We classify a policy as exploitable if one of the vulnerable domains detected by RDScan is whitelisted and the attacker acquires the relevant capabilities to perform the attack, which vary depending on the threat under analysis (see §\ref{subsubsec:attacks-csp}).
For instance, the attacker needs the \capab{file} capability to perform script injection since she needs to upload the payload on a file on the server she controls. Moreover, if the page to attack is served over HTTPS, the \capab{https} capability is required due to the restrictions imposed by browsers on mixed contents~\cite{MixedContent}.

\subsubsection{Analysis of CORS}

To evaluate the security of the CORS policy implemented by a website, we perform multiple requests with different \texttt{Origin} values and inspect the HTTP headers in the response to understand whether CORS has been enabled by the server.
Inspired by the classes of CORS misconfigurations identified in~\cite{ChenJDWCPY18}, we test randomly-generated origins with the following characteristics:
\begin{enumerate*}[label={(\roman*)}]
	\item the domain is a related-domain of the target URL;
	\item the domain starts with the registrable domain of the target URL;
	\item the domain ends with the registrable domain of the target URL.
\end{enumerate*}
While the first test verifies whether CORS is enabled for a related-domain, the other two detect common server-side validation mistakes made by websites who want to enable CORS for their related-domains. Such errors include the search of the registrable domain as a substring or a suffix of the \texttt{Origin} header value, which results in having, e.g., \domain{www.foo.com} whitelisting not only \texttt{a.foo.com} but also \texttt{atkfoo.com}.
For every test, we check:
\begin{itemize}[noitemsep]
	\item whether CORS is enabled for requests without credentials, if the \texttt{Access-Control-Allow-Origin} header is present in the response and its value is that of the \texttt{Origin} header contained in the request;
	\item if the \texttt{Access-Control-Allow-Credentials} header is also present and set to \texttt{true}, which enables CORS for requests with credentials.
\end{itemize}
We report a CORS deployment as vulnerable to web attackers if the second or the third test succeeds. Instead, we report a page as exploitable exclusively by a related-domain attacker if only the first test succeeds and, among the vulnerable related-domains discovered by RDScan, one grants the \capab{js} capability to the attacker.

\subsubsection{Analysis of postMessage Handlers}

PMForce~\cite{SteffensS20} is an automated in-browser framework for the analysis of postMessage event handlers.
It combines selective force execution and taint tracking to extract the constraints on the message contents (e.g., presence of a certain string in the message) that lead to execution traces in which the message enters a dangerous sink that  allows for code execution (e.g., \texttt{eval}) or the alteration of the browser state (e.g., \texttt{document.cookie}).
A message satisfying the extracted constraints is generated using the Z3 solver and the handler under analysis is invoked with the message as parameter to ensure that the exploit is successfully executed.

We integrated PMForce in our pipeline and modified it to generate, for each handler, multiple exploit messages with the same contents but a different \texttt{origin} property, e.g., a related-domain origin and a randomly-generated cross-site origin.
We identify a page as vulnerable to any web attacker if any of its handlers is exploitable from a cross-site position. Instead, we say that a page is exploitable by a \rda{} if its handlers can be exploited only from a related-domain position and one of the vulnerable domains discovered by RDScan grants to the attacker the \capab{js} capability, which is required to open a new tab and send messages to it.

\subsubsection{Analysis of Domain Relaxation}
As a first step, we need to figure out whether the page set the \texttt{document.domain}  property after being loaded. This is easy to spot if the value of the property differs from the domain name of the page; otherwise, it is still possible that the page set \texttt{document.domain} to its original value to enable the relaxation mechanism, as discussed in §\ref{subsec:domain-relaxation}.
To identify this case, we leverage puppeteer APIs to:
\begin{itemize}[noitemsep]
	\item inject a frame from a (randomly generated) subdomain of the page under analysis;
	\item intercept the outgoing network request and provide as response a page with a script that performs domain relaxation and tries to access the DOM of the parent frame.
\end{itemize}
The operation succeeds only if the page has set the \texttt{document.domain} property. In this case, we say that the domain relaxation mechanism is exploitable by a related-domain attacker if RDScan discovered a vulnerable subdomain that grants to the attacker the \capab{js} capability.

\subsection{Limitations}
We acknowledge a number of technical limitations in our approach. The three main components of the pipeline described in Figure~\ref{fig:pipeline} have been executed at different points in time. Having the DNS data collection running first, the subdomain takeover module processed a fraction of subdomains that were no longer resolvable and, similarly, it missed new subdomains that were issued after the completion of the DNS enumeration. This leads to a possible \emph{underestimation} of the threats in the wild concerning unresolvable domains and expired services. The detection of subdomains pointing to potentially deprovisioned cloud instances relies instead on a heuristics which might introduce false positives, see Appendix~\ref{subsubsec:cloud-instances}. 
To avoid confusion, we account for this limitation in the paper by referring to those domains as \emph{potentially vulnerable}.

In general, we made a conscious use of heuristics when we considered them to bring value to the analysis. For instance, we used a simple heuristics to identify cookies which likely contain session identifiers and are thus particularly interesting from a security perspective. Note however that we did not authenticate to the sites under analysis, which is a standard limitation of web security measurements.

\section{Security Evaluation}
\label{sec:results}
We report on the results of our security evaluation on the top 50k domains from the Tranco list. We quantify the vulnerabilities that allow an attacker to be in a related-domain position and we provide a characterization of the affected websites. We review the capabilities and security pitfalls of \dataTotServices{} service providers and discuss the issues affecting institutions using eduroam. We present the outcome of our web analysis and we identify practical vulnerabilities by intersecting the capabilities on vulnerable domains with the threats found on web applications hosted on their related domains.

\subsection{Attack Vectors and Capabilities}
\label{subsec:attack-vectors-and-caps}
RDScan identified \dataVSubdomains{} subdomains exposed to a takeover vulnerability, distributed among \dataVDomains{} domains from the top 50k of the Tranco list. Most of the vulnerabilities are caused by discontinued third-party services (83\%), with expired domains being responsible for the remaining 17\%. The analysis of deprovisioned cloud instances reported \dataVcDomains{} potentially vulnerable domains, confirming the prevalence of this threat as reported in previous work~\cite{LiuHW16}. A breakdown of the attack vectors found in the wild is presented in Table~\ref{tab:attack-vectors}.

\begin{table}[ht]
    \footnotesize
	\caption{Distribution of attack vectors.}
	\label{tab:attack-vectors}
	\begin{tabularx}{\linewidth}{@{}lRR@{}}
	\toprule
	Attack vector &  Domains & Sites \\ \midrule
    Expired domains & \dataVdSubdomains{} & \dataVdDomains{} \\
    Discontinued services & \dataVsSubdomains{} & \dataVsDomains{} \\
    Deprovisioned cloud instances & \dataVcSubdomains{} & \dataVcDomains{} \\
    \textbf{Vulnerable} & \textbf{\dataVSubdomains{}} & \textbf{\dataVDomains{}} \\
    \textbf{Vulnerable + potentially vulnerable} & \textbf{\dataPSubdomains{}} & \textbf{\dataPDomains{}} \\
    \bottomrule
    \end{tabularx}
\end{table}

\subsubsection{Characterization of Vulnerable Domains}
\label{subsubsec:characterization}

\begin{figure*}[ht!]
     \centering
     \begin{subfigure}[b]{0.33\textwidth}
        \centering
        \includegraphics[width=\textwidth]{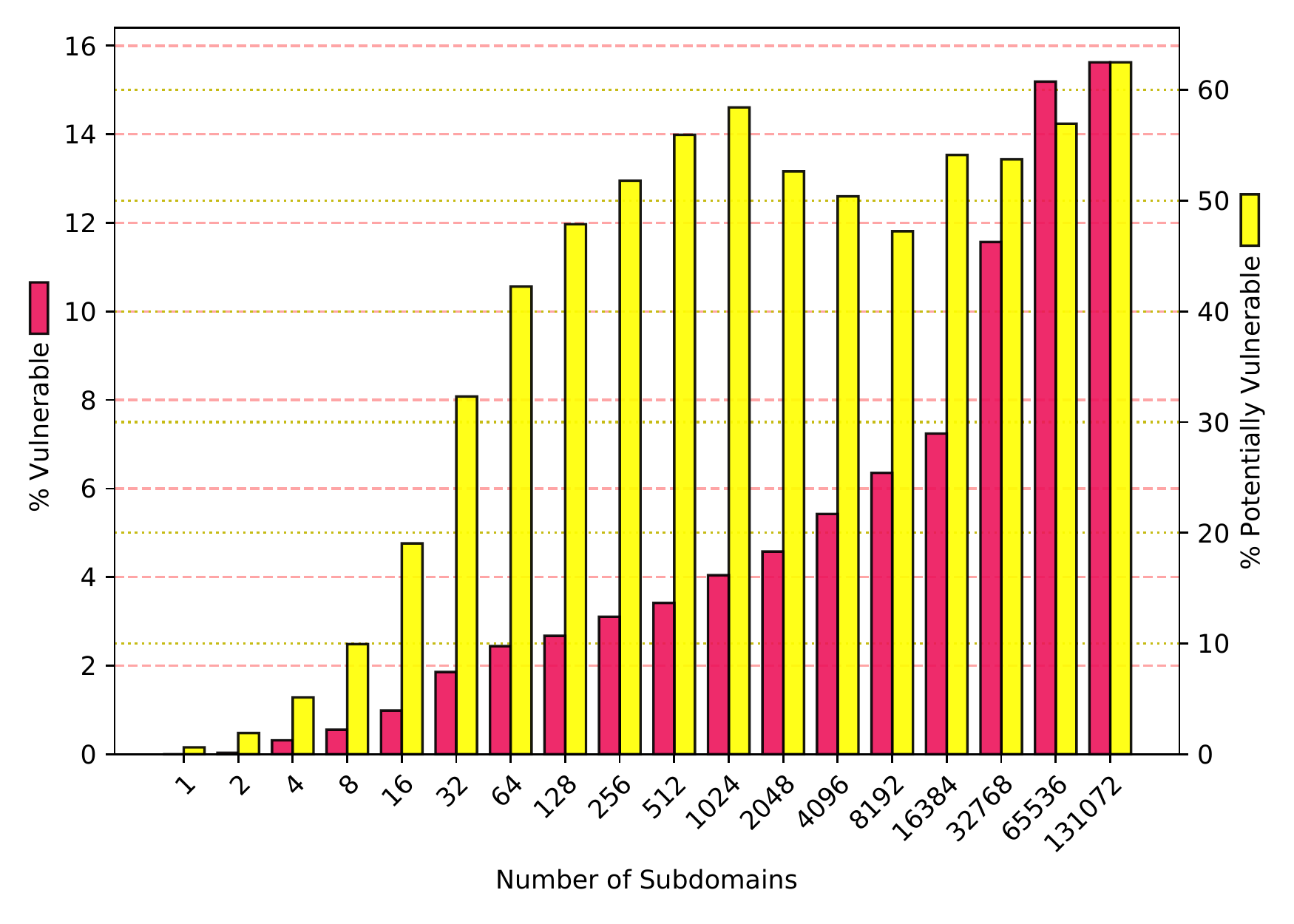}
         \begin{minipage}{.1cm}
           \vfill
         \end{minipage}
        \caption{\# Subdomains}
        \label{fig:characterization:subdomains}
    \end{subfigure}
    \hfill
    \begin{subfigure}[b]{0.33\textwidth}
        \centering
        \includegraphics[width=\textwidth]{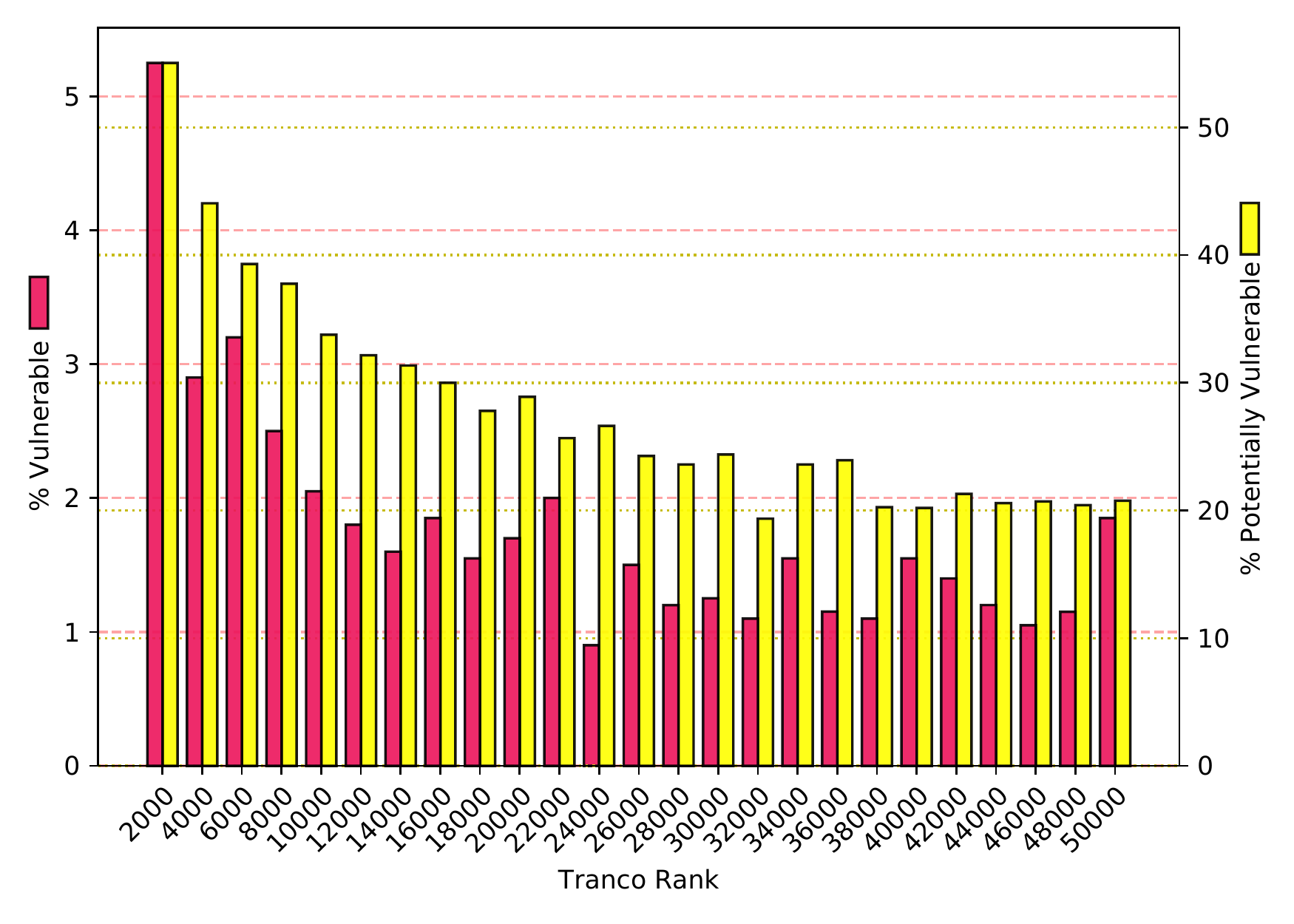}
         \begin{minipage}{.1cm}
           \vfill
         \end{minipage}
        \caption{Tranco Rank}
        \label{fig:characterization:tranco}
    \end{subfigure}
    \hfill
     \begin{subfigure}[b]{0.33\textwidth}
         \centering
         \includegraphics[width=\textwidth]{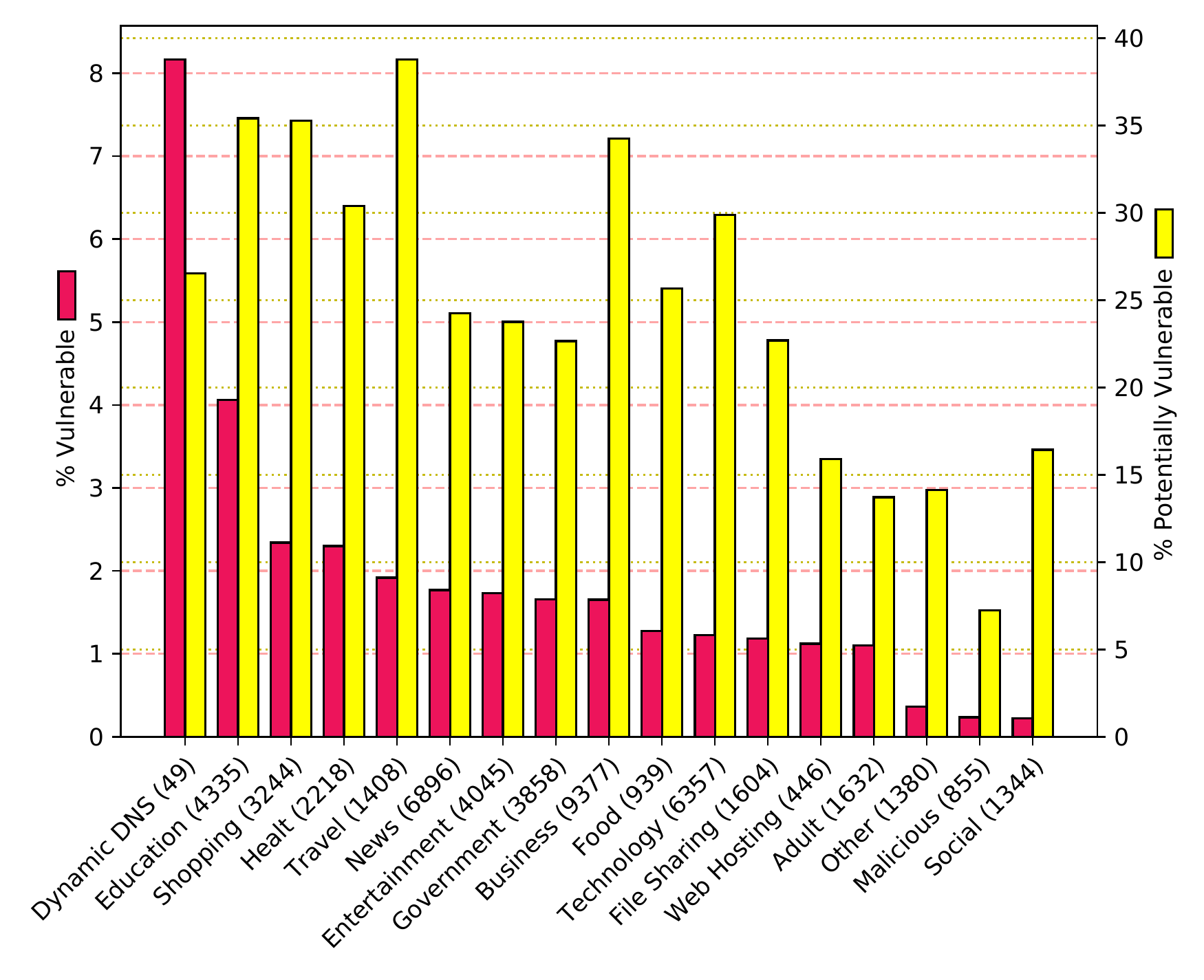}
         \caption{Categories}
         \label{fig:characterization:categories}
     \end{subfigure}
        \caption{Characterization of Vulnerable Domains.}
        \label{fig:characterization}
\end{figure*}

As expected, the likelihood of a domain to be vulnerable is directly related to the breadth of its attack surface, i.e., the number of subdomains we found. Figure~\ref{fig:characterization:subdomains} pictures well this correlation, showing that around 15\% of the domains with more than 50,000 subdomains are vulnerable. A similar consideration applies to potentially vulnerable domains due to the presence of deprovisioned cloud instances.

Figure~\ref{fig:characterization:tranco} outlines the distribution of vulnerable domains depending on the rank of each site in the Tranco list. Sites in top positions are more likely to have a vulnerable subdomain than those with a lower rank. The intuition here is that top sites belong to large organizations with many subdomains.

The analyzed websites have been further partitioned into categories in Figure~\ref{fig:characterization:categories}. Special care has to be taken when considering dynamic DNS: the 49 domains listed in this category are those used by dynamic DNS services, such as \domain{ddns.net}, \domain{noip.com}, \domain{afraid.org}. RDScan identified vulnerable subdomains belonging to 8 domains, but 4 of them resulted to be listed in the PSL. We excluded these domains from our analysis, given that taking control of one of their subdomains would not put the attacker in a related-domain position with respect to the parent domain. The same criteria has been adopted when evaluating service and hosting providers offering subdomains to their users. We refer to §\ref{subsubsec:analysis-services} for a detailed analysis of Dynamic DNS services and hosting providers.

The second most affected category concerns education websites. We found that academic institutions generally have complex and heterogeneous public facing IT infrastructures that translate into a high number of subdomains. By restricting the analysis to the \texttt{.edu} TLD, we observed \dataEduTotSites{} domains having on average \dataEduAvgSubdomains{} subdomains each. The percentage of domains with at least one vulnerable subdomain is \dataEduVulnPercentage{}, which is substantially higher than any other TLD considered. For comparison, the percentage in \texttt{.com} is \dataComVulnPercentage{}.

Overall, we identified vulnerabilities affecting top domains across all the categories. To exemplify, we found subdomain takeover vulnerabilities on news websites like \domain{cnn.com} and \domain{time.com}, university portals like \domain{harvard.edu} and \domain{mit.edu}, governmental websites like \domain{europa.eu} and \domain{nih.gov}, and IT companies like \domain{lenovo.com} and \domain{cisco.com}. Although most of the discovered issues could be easily fixed by routinely checking the validity of DNS records, our large-scale vulnerability assessment raises concerns due to the number and pervasiveness of the identified threats.

\subsubsection{Analysis of Third-Party Services}
\label{subsubsec:analysis-services}

In this work we examined \dataTotThirdPartyServices{} service and hosting providers and \dataTotDynDNS{} dynamic DNS services for a total of \dataTotServices{} third-party services. Our selection comprises services mentioned in previous work~\cite{LiuHW16} and community efforts~\cite{edoverflowtakeover}, excluding those that required a payment to carry out our analysis.

We combined manual testing and review of the documentation to assess the capabilities available to a registered user of each service. We also evaluated the considered services against the security pitfalls described in §\ref{subsubsec:dangling-dns-records}: \begin{enumerate*}[label={(\roman*)}]
    \item if the service allows users to create a website under a specific subdomain, we checked whether the parent domain of the assigned website is in the PSL;
    \item if the domain ownership verification allows attackers to claim subdomains of an already mapped domain (\eg{} due to the presence of a wildcard DNS entries);
    \item if the \texttt{www} subdomain of a mapped domain automatically redirects to the parent domain, \eg{} \domain{www.shop.example.com} redirects to \domain{shop.example.com}, we checked whether \domain{www.shop.example.com} can be claimed from a different account.
\end{enumerate*}
Due to space constraints, we present the full results of our analysis in Table~\ref{tab:service-capabilities} in Appendix~\ref{appendix:services}.

\begin{figure}[t]
    \centering
    \includegraphics[width=0.38\textwidth]{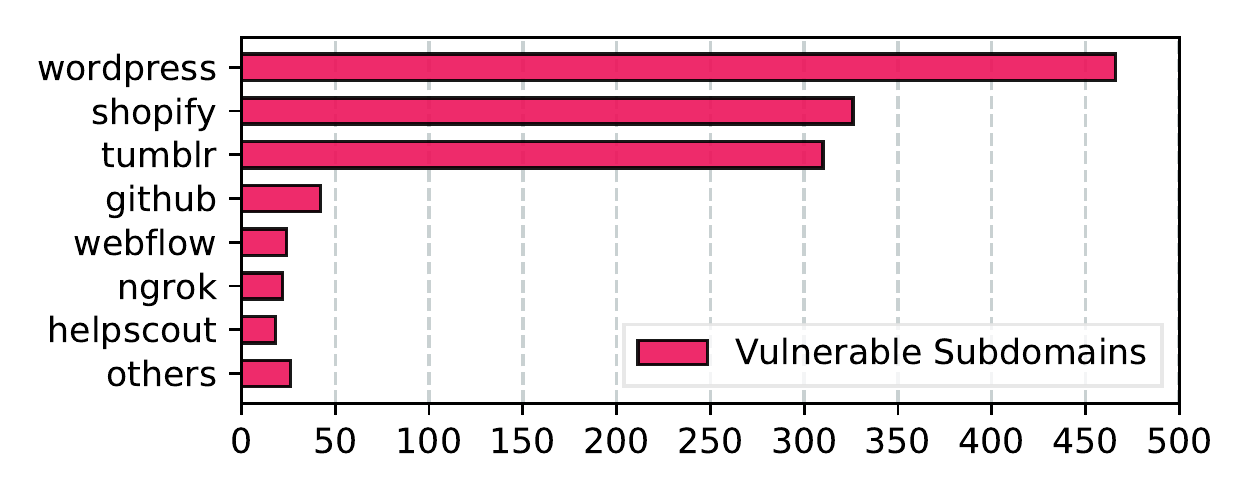}
    \caption{Vulnerable subdomains on discontinued services.}
    \label{fig:services:subdomains}
\end{figure}

\noindent Figure~\ref{fig:services:subdomains} shows the distribution of the vulnerable subdomains across service providers. The majority of the vulnerable subdomains (93\%) are hosted on the first four most used services: Wordpress, Shopify, Tumblr and Github Pages. These prominent services give users the ability to host a website with a valid TLS certificate for the associated domain. Users are allowed to customize the markup and JavaScript code of the pages, and for Tumblr and GitHub Pages, users are allowed to upload arbitrary files in their websites. In general, the capabilities obtained by an attacker controlling a service vary depending on the specific platform, ranging from \capab{content} only (UptimeRobot), to full host control (ngrok). We found that 19 out of \dataTotThirdPartyServices{} services grant the \capab{js} and \capab{https} capabilities, while 21 provide the \capab{js} capability alone. The \capab{file} capability is the most uncommon, being available only on 4 services.

Surprisingly, we discovered that in 20 out of the \dataTotServices{} analyzed services, any registered user controls a website that is in a related-domain position to all the other websites hosted on the platform. Tumblr and Wordpress along with 11 other services even share their primary domain with user controlled websites, \eg{}, \domain{attacker.tumblr.com} is related to \domain{tumblr.com}. Only GitHub and ngrok prevent this threat by including the apex domains assigned to their users in the PSL.

Lastly, we found that 17 out of the 21 services we could test, have issues in the ownership verification mechanism. Among the four most used services, only Wordpress prevents attackers to claim subdomains of an already mapped domains. Moreover, 8 service providers perform an automatic redirection from the \texttt{www} subdomain to the parent domain. Only Shopify and Launchrock do not prevent this subdomain from being mapped to different accounts. Users of these service might thus assume that the \texttt{www} subdomain is implicitly bound to their account and cannot be claimed by others.

Given the prevalence of the aforementioned issues, in addition to our large-scale disclosure campaign (see Appendix~\ref{appendix:disclosure}), we reported to GitHub and Shopify -- two of the major service providers -- the vulnerabilities found in their domain ownership verification process. GitHub acknowledged the vulnerability and told us that they ``[...] are exploring various changes to the custom domain flow that will improve this situation by requiring formal domain ownership verification''. Shopify recognized the importance of the reported issue, and despite being out of scope for their bug bounty program, they decided to award us with \$1,000. Specifically, we noticed that some merchants relied on the www-prefixed version of their domain, without an explicit association of this domain to their account. Attackers claiming the www-prefixed domain would have a much higher gain than being in a related-domain position, since they control the main entry point of the e-shop.

\paragraph{Dynamic DNS Services.}%
We observed that only 2 providers listed all their domains in the PSL. Noip and DynDNS left out a small number of the domains they offer, but it is not clear to us whether this is due to negligence or if this is a deliberate choice. Instead, FreeDNS, with more 50k domains, did not include any of them in the list, leaving their massive user base at risk. We reported this major flaw to the FreeDNS maintainer who acknowledged it but took no action, as it would be impossible to maintain an updated list of thousands of domains in the PSL, given the lack of an API to manage PSL entries.
We refer the interested reader to Table~\ref{tab:dyndns-psl} in Appendix~\ref{appendix:services} for additional details.

\subsection{Web Threats}
\label{subsec:webthreats}
We now turn the attention to the web application security implications of our analysis. We start by discussing cookie confidentiality and integrity, as summarized in Table~\ref{tab:cookies}. 

\begin{table}[t]
    \footnotesize
    \caption{Results of the cookie security analysis.}
	\label{tab:cookies}
    \begin{tabularx}{\linewidth}{@{}lXXX@{}}
		\toprule
		\textbf{Cookies} & \textbf{Count} & \textbf{Domains} & \textbf{Sites} \\
		\midrule
		Total number & \dataCountCookies & \dataOriginsCookies & \dataDomainsCookies \\
		Domain cookies & \dataCountSessionCookies & \dataOriginsSessionCookies & \dataDomainsSessionCookies \\
		Confidentiality issues & \dataCountCookiesConfidentiality & \dataOriginsCookiesConfidentiality & \dataDomainsCookiesConfidentiality \\
		Integrity issues & \dataCountCookiesIntegrity & \dataOriginsCookiesIntegrity & \dataDomainsCookiesIntegrity \\
		\bottomrule
	\end{tabularx}
\end{table}

Overall, our crawler collected \dataCountCookies{} cookies flagged as session cookies by our heuristics, including \dataCountSessionCookies{} domain cookies. Out of these, we identify \dataCountCookiesConfidentiality{} (14\%) cookies from \dataDomainsCookiesConfidentiality{} (81\%) sites whose confidentiality can be violated by a related-domain attacker. This shows that related-domain attackers can often get access to session cookies, which may enable attacks like session hijacking. Our analysis also shows that the state of cookie integrity is even worse: in particular, we identify \dataCountCookiesIntegrity{} (99\%) session cookies from \dataDomainsCookiesIntegrity{} (99\%) sites which do not have integrity against a related-domain attacker, hence may enable attacks like session fixation and cookie forcing. This increase comes from the fact that related-domain attackers can compromise the confidentiality of domain cookies alone, while they can break the integrity of any cookie by exploiting \emph{cookie shadowing}~\cite{ZhengJLDCWW15}. The only robust way to improve cookie integrity in this setting is the adoption of the \texttt{\_\_Host-} prefix, which is unfortunately negligible in the wild: we only identified one cookie using it in our dataset.

Table~\ref{tab:web-all} presents the analysis results for the other points considered in our study. We start by discussing the insights for CSP. The first observation we make is that, as observed by previous studies~\cite{WeichselbaumSLJ16,CalzavaraRB18,RothBCNS20}, the majority of CSPs in the wild suffers from incorrect configurations, voiding their security guarantees even against web attackers. Remarkably, however, related-domain attackers are more powerful than traditional web attackers for real-world CSPs. In particular, this is apparent for object injection, frame injection and framing control. For example, we quantified the following increase in the attack surface for frame injection: \dataOriginsCSPFramePotentiallyVulnerableReldom{} (+21\%) new potentially exploitable cases, out of which \dataOriginsCSPFrameVulnerableReldom{} (+7\%) are actually exploitable by controlling one of the vulnerable subdomains identified in our dataset.

As to the other mechanisms, CORS deployments are significantly more at risk against related-domain attackers rather than against traditional web attackers. In particular, we identify \dataOriginsCORSVulnerableReldom{} (+13\%) new exploitable cases, including \dataOriginsCORSCredsVulnerableReldom{} (+54\%) cases with credentials. Note that the use of CORS with credentials is particularly delicate from a security perspective, hence the strong percentage increase in the number of vulnerable cases is concerning. Domain relaxation, instead, can be abused by related-domain attackers in \dataOriginsRelaxationVulnerableReldom{} domains over \dataDomainsRelaxationVulnerableReldom{} sites. Though these numbers are relatively low, their importance should be analyzed with care. First, exploiting domain relaxation puts a related-domain attacker in the same origin of the target web application, hence bypassing all web security boundaries: this is a critical vulnerability, which deserves attention. Second, our analysis results are limited to the number of subdomains which we actually managed to control, hence the actual room for exploitation may be larger: domain relaxation is a bad security practice, which should better be avoided on the modern Web. Finally, our analysis of postMessage shows that all sites suffering from unsafe programming practices are already vulnerable against web attackers, i.e., for this specific attack vector related-domain attackers are no more powerful than traditional web attackers, at least based on the collected data. In other words, sites either do not enforce any security check or restrict communication to selected individual origins: this might be a consequence of the postMessage API granting access to origin information, rather than site information directly.

\begin{table*}
	\centering
    \footnotesize
   	\caption{Web application security abuses by related-domain attackers.}
	\label{tab:web-all}\vspace{-5pt}
	\begin{tabular}{lcccccccc}
		\toprule
		\multirow{3}{*}{\textbf{Mechanism}} & \multicolumn{2}{c}{\multirow{2}{*}{\textbf{Deployed}}} & \multicolumn{2}{c}{\textbf{Exploitable by}} & \multicolumn{2}{c}{\textbf{Potentially Exploitable}} & \multicolumn{2}{c}{\multirow{2}{*}{\textbf{Exploitable}}} \\
		&&& \multicolumn{2}{c}{\textbf{any Web Attacker}} & \multicolumn{2}{c}{\textbf{by Related-Domain Attackers}} \\
		& \textbf{Domains} & \textbf{Sites} & \textbf{Domains} & \textbf{Sites} & \textbf{Domains} & \textbf{Sites} & \textbf{Domains} & \textbf{Sites} \\
		\midrule
		CSP: script & \dataOriginsCSPScript & \dataDomainsCSPScript & \dataOriginsCSPScriptVulnerableWeb & \dataDomainsCSPScriptVulnerableWeb  & \dataOriginsCSPScriptPotentiallyVulnerableReldom & \dataDomainsCSPScriptPotentiallyVulnerableReldom & \dataOriginsCSPScriptVulnerableReldom & \dataDomainsCSPScriptVulnerableReldom \\
		CSP: style & \dataOriginsCSPStyle & \dataDomainsCSPStyle & \dataOriginsCSPStyleVulnerableWeb & \dataDomainsCSPStyleVulnerableWeb & \dataOriginsCSPStylePotentiallyVulnerableReldom & \dataDomainsCSPStylePotentiallyVulnerableReldom & \dataOriginsCSPStyleVulnerableReldom & \dataDomainsCSPStyleVulnerableReldom \\
		CSP: object & \dataOriginsCSPObject & \dataDomainsCSPObject & \dataOriginsCSPObjectVulnerableWeb & \dataDomainsCSPObjectVulnerableWeb & \dataOriginsCSPObjectPotentiallyVulnerableReldom & \dataDomainsCSPObjectPotentiallyVulnerableReldom & \dataOriginsCSPObjectVulnerableReldom & \dataDomainsCSPObjectVulnerableReldom \\
		CSP: frame & \dataOriginsCSPFrame & \dataDomainsCSPFrame & \dataOriginsCSPFrameVulnerableWeb & \dataDomainsCSPFrameVulnerableWeb & \dataOriginsCSPFramePotentiallyVulnerableReldom & \dataDomainsCSPFramePotentiallyVulnerableReldom & \dataOriginsCSPFrameVulnerableReldom & \dataDomainsCSPFrameVulnerableReldom \\
		CSP: framing control & \dataOriginsCSPFraming & \dataDomainsCSPFraming & \dataOriginsCSPFramingVulnerableWeb & \dataDomainsCSPFramingVulnerableWeb  & \dataOriginsCSPFramingPotentiallyVulnerableReldom & \dataDomainsCSPFramingPotentiallyVulnerableReldom & \dataOriginsCSPFramingVulnerableReldom & \dataDomainsCSPFramingVulnerableReldom \\
		CORS: total cases & - & - & \dataOriginsCORSVulnerableWeb & \dataDomainsCORSVulnerableWeb & - & - & \dataOriginsCORSVulnerableReldom & \dataDomainsCORSVulnerableReldom \\
		CORS: with credentials & - & - & \dataOriginsCORSCredsVulnerableWeb & \dataDomainsCORSCredsVulnerableWeb & - & - & \dataOriginsCORSCredsVulnerableReldom & \dataDomainsCORSCredsVulnerableReldom \\
		postMessage & \dataOriginsPostMessage & \dataDomainsPostMessage & \dataOriginsPostMessageVulnerableWeb & \dataDomainsPostMessageVulnerableWeb & - & - & \dataOriginsPostMessageVulnerableReldom & \dataDomainsPostMessageVulnerableReldom \\
		Domain Relaxation & \dataOriginsRelaxation & \dataDomainsRelaxation & 0 & 0 & - & - & \dataOriginsRelaxationVulnerableReldom & \dataDomainsRelaxationVulnerableReldom \\
		\bottomrule
	\end{tabular}
\end{table*}

\section{Related Work}
\label{sec:related}
\paragraph{Related-Domain Attackers.}
The notion of related-domain attacker was first introduced by Bortz, Barth and Czeskis~\cite{Bortz11}. Their work identified the security risks posed by related domains against (session) cookies and proposed a possible solution called \emph{origin cookies}. A similar defense mechanism, i.e., the \texttt{\_\_Host-} prefix, was eventually integrated in major web browsers. Other than that, related-domain attackers received only marginal attention by the security community, with a few notable exceptions. Zheng et al. discussed the security implications of the lack of cookie integrity in many top sites, considering both network and related-domain attackers~\cite{ZhengJLDCWW15}. Calzavara et al. presented black-box testing strategies for web session integrity, including related-domain attackers in their threat model~\cite{CalzavaraRRB19}. Related-domain attackers have also been considered in formal web security models, again in the context of web sessions~\cite{CalzavaraFGMT20}. Our paper significantly advances the understanding of related-domain attackers by discussing new security threats, which go beyond web sessions and have been quantified in the wild through a large-scale measurement.

\paragraph{Attacking Subdomains.}
Subdomain takeover is an infamous attack, which has been covered by a body of work. Liu \etal~\cite{LiuHW16} studied the threat posed by dangling DNS records, \eg{}, records that contain aliases to expired domains or pointing to IP addresses hosted on cloud services. The authors performed a large-scale analysis that uncovered the existence of hundreds of dangling records among the subdomains of the top 10k sites of Alexa and under the \texttt{.edu} and \texttt{.gov} zones.
With respect to~\cite{LiuHW16}, we improved the subdomain enumeration part by a factor of 13 and increased the number of analyzed services from 9 to \dataTotServices. Also, the paper does not extensively analyze the web security implications of subdomain takeover.
Instead, Borgolte \etal~\cite{BorgolteFHKV18} improved on the results of~\cite{LiuHW16} concerning deprovisioned cloud instances and proposed an extension of the ACME protocol~\cite{RFC8555} used by some CAs for domain validation (\eg, \emph{Let's Encrypt}).
Liu \etal~\cite{LiuLDWLD17} proposed a technique to detect shadowed domains used in malware distribution campaigns, \ie{}, legitimate domains that are compromised to spawn an arbitrary number of subdomains after taking control of the DNS configuration panel at the registrar. Alowaisheq \etal~\cite{Alowaisheq20} recently demonstrated a domain hijacking attack that relies on the exploitation of stale \texttt{NS} records. We consider these threats out of the scope of our analysis, as they have different security implications and are less widespread than the vulnerabilities we discuss.

\paragraph{Web Measurements.}
Chen \etal~\cite{ChenJDWCPY18} performed a large-scale measurement of CORS misconfigurations. Among the 480k domains that they analyzed, they discovered that 27.5\% of them are affected by some vulnerability and, in particular, 84k trust all their subdomains and can thus be exploited by a related-domain attacker. Son and Shmatikov~\cite{SonS13} analyzed the usage of the Messaging API on the top 10k Alexa websites. The authors found that 1.5k hosts do not perform any origin checking on the receiving message and 261 implement an incorrect check: (almost) all these checks can be bypassed from a related-domain position, although half of them can also be bypassed from domains with a specially-crafted name.
In more recent work, Steffens and Stock~\cite{SteffensS20} proposed an automated framework for the analysis of postMessage handlers and used it perform a comprehensive analysis of the first top 100k websites of the Tranco list. The authors  discovered 111 vulnerable handlers, out of which 80 do not perform any origin check. Regarding the remaining handlers, the authors identified only 8 incorrect origin validations, showing an opposite trend with respect to~\cite{SonS13}.
Finally, insecure configurations of CSP have been analyzed in a number of research papers~\cite{WeissbacherLR14,WeichselbaumSLJ16,CalzavaraRB18,RothBCNS20}. However, none of these works considered the problem of related-domain attacks.

\section{Conclusion}
\label{sec:conclusion}
In this paper we presented the first analysis tailored at quantifying the threats posed by related-domain attackers to the security of web applications. We first introduced a novel framework that captures the capabilities acquired by such attackers, according to the position in which they operate, and we discuss which web attacks can be launched from that privileged position, highlighting the advantages with respect to traditional web attackers. We also studied the security implications of \dataTotServices{} third-party service providers and dynamic DNS to identify the capabilities that a \rda{} acquires when taking over a domain hosted by them, and present a novel subdomain hijacking technique that resulted in a bug bounty of \$1,000. Next we presented our automated toolchain to assess the pervasiveness of these threats in the wild. The toolchain consists of an analysis module for subdomain takeover that identifies which subdomains can be hijacked by an attacker. Next, the web security module quantifies how many related-domains can be attacked from the domains discovered in the previous step. We performed a large-scale analysis on the 50k most popular domains and we identified  vulnerabilities in \dataVDomains{} of them, including major websites like \domain{cnn.com} and \domain{cisco.com}. Then, we correlated for the first time the impact of these vulnerabilities on the security of web applications, showing that related-domain attackers have an additional gain compared to web attackers that goes beyond the traditional cookie issues.

\paragraph{Acknowledgments.}
We thank Google for sponsoring our research with \$5,000 in credits for Google Cloud Platform and Cisco Talos for granting us access to a dataset that was used during a preliminary investigation for this project.

\bibliographystyle{plain}
\bibliography{biblio}

\begin{thebibliography}{10}

\bibitem{Puppeteer}
Puppeteer.
\newblock \url{https://pptr.dev/}, 2020.

\bibitem{AasEtAl19}
Josh Aas, Richard Barnes, Benton Case, Zakir Durumeric, Peter Eckersley, Alan
  Flores-L\'{o}pez, J.~Alex Halderman, Jacob Hoffman-Andrews, James Kasten,
  Eric Rescorla, Seth Schoen, and Brad Warren.
\newblock {Let's Encrypt: An Automated Certificate Authority to Encrypt the
  Entire Web}.
\newblock In {\em Proceedings of the 2019 ACM SIGSAC Conference on Computer and
  Communications Security, {CCS} 2019}, 2019.

\bibitem{Abusix}
{Abusix}.
\newblock {Abuse Contact Database}.
\newblock \url{https://www.abusix.com/contactdb}, 2020.

\bibitem{Alowaisheq20}
Eihal Alowaisheq, Siyuan Tang, Zhihao Wang, Fatemah Alharbi, Xiaojing Liao, and
  XiaoFeng Wang.
\newblock {Zombie Awakening: Stealthy Hijacking of Active Domains Through DNS
  Hosting Referral}.
\newblock In {\em Proceedings of the 2020 {ACM} {SIGSAC} Conference on Computer
  and Communications Security, {CCS} 2020}, 2020.

\bibitem{RFC8555}
Richard Barnes, Jacob Hoffman-Andrews, Daniel McCarney, and James Kasten.
\newblock {Automatic Certificate Management Environment (ACME)}.
\newblock \url{https://tools.ietf.org/html/rfc8555}, March 2019.

\bibitem{Barth08}
Adam Barth, Collin Jackson, and John~C. Mitchell.
\newblock Robust defenses for cross-site request forgery.
\newblock In {\em Proceedings of the 2008 {ACM} Conference on Computer and
  Communications Security, {CCS} 2008, Alexandria, Virginia, USA, October
  27-31, 2008}. {ACM}, 2008.

\bibitem{Biasini15}
Nick Biasini.
\newblock {Threat Spotlight: Angler Lurking in the Domain Shadows}.
\newblock \url{http://blogs.cisco.com/security/talos/angler-domain-shadowing},
  2015.

\bibitem{RFC5280}
Sharon Boeyen, Stefan Santesson, Tim Polk, Russ Housley, Stephen Farrell, and
  Dave Cooper.
\newblock {Internet X.509 Public Key Infrastructure Certificate and Certificate
  Revocation List (CRL) Profile}.
\newblock \url{https://tools.ietf.org/html/rfc5280}, May 2008.

\bibitem{BorgolteFHKV18}
Kevin Borgolte, Tobias Fiebig, Shuang Hao, Christopher Kruegel, and Giovanni
  Vigna.
\newblock {Cloud Strife: Mitigating the Security Risks of Domain-Validated
  Certificates}.
\newblock In {\em Proceedings of the 25th Annual Network and Distributed System
  Security Symposium, {NDSS} 2018}, 2018.

\bibitem{Bortz11}
Andrew Bortz, Adam Barth, and Alexei Czeskis.
\newblock {Origin Cookies: Session Integrity for Web Applications}.
\newblock In {\em Web 2.0 Security \& Privacy Workshop {W2SP 2011}}, 2011.

\bibitem{BrinP98}
Sergey Brin and Lawrence Page.
\newblock The anatomy of a large-scale hypertextual web search engine.
\newblock {\em Comput. Networks}, 30(1-7):107--117, 1998.

\bibitem{Bugcrowd}
{Bugcrowd}.
\newblock {Public Bug Bounty List}.
\newblock \url{https://www.bugcrowd.com/bug-bounty-list/}, 2020.

\bibitem{BugliesiCFK15}
Michele Bugliesi, Stefano Calzavara, Riccardo Focardi, and Wilayat Khan.
\newblock Cookiext: Patching the browser against session hijacking attacks.
\newblock {\em Journal of Computer Security}, 23(4):509--537, 2015.

\bibitem{CalzavaraFGMT20}
Stefano Calzavara, Riccardo Focardi, Niklas Grimm, Matteo Maffei, and Mauro
  Tempesta.
\newblock Language-based web session integrity.
\newblock In {\em Proceedings of the 33rd {IEEE} Computer Security Foundations
  Symposium, {CSF} 2020}, pages 107--122, 2020.

\bibitem{CalzavaraRB18}
Stefano Calzavara, Alvise Rabitti, and Michele Bugliesi.
\newblock Semantics-based analysis of content security policy deployment.
\newblock {\em {ACM} Transactions on the Web}, 12(2):10:1--10:36, 2018.

\bibitem{CalzavaraRRB19}
Stefano Calzavara, Alvise Rabitti, Alessio Ragazzo, and Michele Bugliesi.
\newblock Testing for integrity flaws in web sessions.
\newblock In {\em Proceedings of the 24th European Symposium on Research in
  Computer Security, {ESORICS} 2019}, pages 606--624, 2019.

\bibitem{Censys}
Censys.
\newblock \url{https://censys.io/}, 2020.

\bibitem{ChenJDWCPY18}
Jianjun Chen, Jian Jiang, Hai{-}Xin Duan, Tao Wan, Shuo Chen, Vern Paxson, and
  Min Yang.
\newblock {We Still Don't Have Secure Cross-Domain Requests: an Empirical Study
  of CORS}.
\newblock In {\em Proceedings of the 27th {USENIX} Security Symposium, {USENIX}
  Security 2018.}, 2018.

\bibitem{CommonCrawl}
Common Crawl.
\newblock Host- and domain-level web graphs feb/mar/may 2020.
\newblock
  \url{https://commoncrawl.org/2020/06/host-and-domain-level-web-graphs-febmarmay-2020/},
  2020.

\bibitem{RFC3912}
Leslie Daigle.
\newblock {WHOIS Protocol Specification}.
\newblock \url{https://tools.ietf.org/html/rfc3912}, September 2004.

\bibitem{edoverflowtakeover}
{EdOverflow}.
\newblock {can-i-take-over-xyz}.
\newblock \url{https://github.com/EdOverflow/can-i-take-over-xyz}.

\bibitem{SecurityTXT}
Edwin Foudil and Yakov Shafranovich.
\newblock {A File Format to Aid in Security Vulnerability Disclosure}.
\newblock \url{https://tools.ietf.org/id/draft-foudil-securitytxt-10.txt},
  August 2020.

\bibitem{FreeDNS}
FreeDNS.
\newblock {Free DNS Hosting, Dynamic DNS Hosting, Static DNS Hosting, subdomain
  and domain hosting}.
\newblock \url{https://freedns.afraid.org/}, 2020.

\bibitem{RFC8659}
Phillip Hallam-Baker, Rob Stradling, and Jacob Hoffman-Andrews.
\newblock {DNS Certification Authority Authorization (CAA) Resource Record}.
\newblock \url{https://tools.ietf.org/html/rfc8659}, November 2019.

\bibitem{CSRFDead}
Scott Helme.
\newblock {Cross-Site Request Forgery is dead!}
\newblock \url{https://scotthelme.co.uk/csrf-is-dead/}, February 2017.

\bibitem{RZD}
IANA.
\newblock Root zone database.
\newblock \url{https://www.iana.org/domains/root/db}.

\bibitem{Kolsek02}
Mitja Kolsek.
\newblock {Session fixation vulnerability in web-based applications}.
\newblock \url{http://www.acrossecurity.com/papers/session_fixation.pdf}, 2002.

\bibitem{RFC4592}
Edward~P. Lewis.
\newblock {The Role of Wildcards in the Domain Name System}.
\newblock \url{https://tools.ietf.org/html/rfc4592}, July 2006.

\bibitem{RFC5936}
Edward~P. Lewis and Alfred Hoenes.
\newblock {DNS Zone Transfer Protocol (AXFR)}.
\newblock \url{https://tools.ietf.org/html/rfc5936}, June 2010.

\bibitem{LiDCKB16}
Frank Li, Zakir Durumeric, Jakub Czyz, Mohammad Karami, Michael Bailey, Damon
  McCoy, Stefan Savage, and Vern Paxson.
\newblock You{\textquoteright}ve got vulnerability: Exploring effective
  vulnerability notifications.
\newblock In {\em Proceedings of the 25th {USENIX} Security Symposium ({USENIX}
  Security 16)}, 2016.

\bibitem{LiuHW16}
Daiping Liu, Shuai Hao, and Haining Wang.
\newblock {All Your DNS Records Point to Us: Understanding the Security Threats
  of Dangling DNS Records}.
\newblock In {\em Proceedings of the 23rd {ACM} Conference on Computer and
  Communications Security, {CCS} 2016}, 2016.

\bibitem{LiuLDWLD17}
Daiping Liu, Zhou Li, Kun Du, Haining Wang, Baojun Liu, and Hai{-}Xin Duan.
\newblock {Don't Let One Rotten Apple Spoil the Whole Barrel: Towards Automated
  Detection of Shadowed Domains}.
\newblock In {\em Proceedings of the 2017 {ACM} {SIGSAC} Conference on Computer
  and Communications Security, {CCS} 2017}, 2017.

\bibitem{GithubCookies}
Vicent Mart\`{i}.
\newblock {Yummy Cookies Across Domains}.
\newblock \url{https://github.blog/2013-04-09-yummy-cookies-across-domains/},
  April 2013.

\bibitem{MozillaSameSite}
{Mike Conca}.
\newblock {Changes to SameSite Cookie Behavior – A Call to Action for Web
  Developers}.
\newblock
  \url{https://hacks.mozilla.org/2020/08/changes-to-samesite-cookie-behavior/},
  August 2020.

\bibitem{RFC1035}
Paul Mockapetris.
\newblock {Domain names - implementation and specification}.
\newblock \url{https://tools.ietf.org/html/rfc1035}, November 1987.

\bibitem{MixedContent}
Mozilla.
\newblock Mixed content.
\newblock
  \url{https://developer.mozilla.org/en-US/docs/Web/Security/Mixed_content}.

\bibitem{PSL}
Mozilla.
\newblock Public suffix list.
\newblock \url{https://publicsuffix.org/}.

\bibitem{RFC8615}
Mark Nottingham.
\newblock {Well-Known Uniform Resource Identifiers (URIs)}.
\newblock \url{https://tools.ietf.org/html/rfc8615}, May 2019.

\bibitem{Zdnet}
Charlie Osborne.
\newblock {Uber patches security flaw leading to subdomain takeover}.
\newblock {ZDNet},
  \url{https://www.zdnet.com/article/uber-patches-security-flaw-leading-to-subdomain-takeover/},
  2017.

\bibitem{Amass}
OWASP.
\newblock Amass.
\newblock \url{https://owasp.org/www-project-amass/}, 2020.

\bibitem{LePochat19}
Victor~Le Pochat, Tom~Van Goethem, Samaneh Tajalizadehkhoob, Maciej
  Korczy\'{n}ski, and Wouter Joosen.
\newblock {Tranco: A Research-Oriented Top Sites Ranking Hardened Against
  Manipulation}.
\newblock In {\em Proceedings of the 26th Annual Network and Distributed System
  Security Symposium}, NDSS 2019, February 2019.

\bibitem{Rapid7}
{Rapid7 Labs}.
\newblock {Open Data, TCP and UDP scans}.
\newblock \url{https://opendata.rapid7.com/}, 2020.

\bibitem{ReisMO19}
Charles Reis, Alexander Moshchuk, and Nasko Oskov.
\newblock {Site Isolation: Process Separation for Web Sites within the
  Browser}.
\newblock In {\em Proceedings of the 28th {USENIX} Security Symposium, {USENIX}
  Security 2019}, 2019.

\bibitem{RothBCNS20}
Sebastian Roth, Timothy Barron, Stefano Calzavara, Nick Nikiforakis, and Ben
  Stock.
\newblock Complex security policy? {A} longitudinal analysis of deployed
  content security policies.
\newblock In {\em Proceedings of the 27th Network and Distributed System
  Security Symposium, {NDSS} 2020}, 2020.

\bibitem{Crtsh}
Sectigo.
\newblock Crt.sh: Certificate search.
\newblock https://crt.sh/, 2020.

\bibitem{SonS13}
Sooel Son and Vitaly Shmatikov.
\newblock {The Postman Always Rings Twice: Attacking and Defending postMessage
  in HTML5 Websites}.
\newblock In {\em Proceedings of the 20th Annual Network and Distributed System
  Security Symposium, {NDSS} 2013}, 2013.

\bibitem{SteffensS20}
Marius Steffens and Ben Stock.
\newblock {PMForce: Systematically Analyzing postMessage Handlers at Scale}.
\newblock In {\em Proceedings of the 27th {ACM} Conference on Computer and
  Communications Security, {CCS} 2020}, 2020.

\bibitem{StockPLBR18}
Ben Stock, Giancarlo Pellegrino, Frank Li, Michael Backes, and Christian
  Rossow.
\newblock {Didn't You Hear Me? - Towards More Successful Web Vulnerability
  Notifications}.
\newblock In {\em 25th Annual Network and Distributed System Security
  Symposium, {NDSS} 2018, San Diego, California, USA, February 18-21, 2018},
  2018.

\bibitem{StockPRJB16}
Ben Stock, Giancarlo Pellegrino, Christian Rossow, Martin Johns, and Michael
  Backes.
\newblock {Hey, You Have a Problem: On the Feasibility of Large-Scale Web
  Vulnerability Notification}.
\newblock In {\em 25th {USENIX} Security Symposium, {USENIX} Security 16,
  Austin, TX, USA, August 10-12, 2016}, 2016.

\bibitem{ChromiumSameSite}
{The Chromium Projects}.
\newblock {SameSite Updates}.
\newblock \url{https://www.chromium.org/updates/same-site}, 2020.

\bibitem{CSPv3}
W3C.
\newblock {Content Security Policy Level 3}.
\newblock \url{https://www.w3.org/TR/CSP3/}, 2018.

\bibitem{WeichselbaumSLJ16}
Lukas Weichselbaum, Michele Spagnuolo, Sebastian Lekies, and Artur Janc.
\newblock {CSP} is dead, long live csp! on the insecurity of whitelists and the
  future of content security policy.
\newblock In {\em Proceedings of the 23rd {ACM} Conference on Computer and
  Communications Security, {CCS} 2016}, pages 1376--1387, 2016.

\bibitem{WeissbacherLR14}
Michael Weissbacher, Tobias Lauinger, and William~K. Robertson.
\newblock Why is {CSP} failing? trends and challenges in {CSP} adoption.
\newblock In {\em Proceedings of the 17th International Symposium on Research
  in Attacks, Intrusions and Defenses, {RAID} 2014}, pages 212--233, 2014.

\bibitem{RFC6265bit06}
Mike West and John Wilander.
\newblock {Cookies: HTTP State Management Mechanism}.
\newblock \url{https://tools.ietf.org/html/draft-ietf-httpbis-rfc6265bis-06},
  April 2020.

\bibitem{ZhengJLDCWW15}
Xiaofeng Zheng, Jian Jiang, Jinjin Liang, Hai{-}Xin Duan, Shuo Chen, Tao Wan,
  and Nicholas Weaver.
\newblock {Cookies Lack Integrity: Real-World Implications}.
\newblock In {\em 24th {USENIX} Security Symposium, {USENIX} Security 15,
  Washington, D.C., USA, August 12-14, 2015}, 2015.

\end{thebibliography}

\appendix
\section{RDScan Internals}
\label{appendix:rdscan}
We provide a detailed overview of the operations performed by RDScan to detect the presence of subdomain takeover vulnerabilities at scale.

\subsection{Expired Domains}
The detection of expired domains follows the procedure described in Algorithm~\ref{alg:expired-domains}. Given a resolving chain that begins with a \texttt{CNAME} record, our tool checks if it points to an unresolvable resource and extracts the eTLD+1 of the canonical name at the end of the chain, that we call $apex$ for brevity. Then, if executing the \texttt{whois} command on the $apex$ domain does not return any match, we check on GoDaddy if the domain is available. In this case, we consider the domain of the resolving chain, i.e., the \emph{alias} of the first record of the chain, as vulnerable.

\begin{algorithm}
	\small
	\caption{Detection of Expired Domains}\label{alg:expired-domains}
	\begin{algorithmic}[1]
	\Require{Set of DNS resolving chains $RC$}
	\Ensure{Set of vulnerable subdomains $V_d$}
	\Procedure{expired\_domains}{$RC$}
		\State $V_d \gets \emptyset$
		\For{\textbf{each} $chain \in RC$}
			\State $head \gets first(chain)$
			\State $tail \gets last(chain)$
			\If{$type(head) = \texttt{CNAME} \:\land\: type(tail) = \texttt{CNAME}$}\label{alg:expired-domains:begins-ends-cname}
				\LineComment{Get the canonical name of the CNAME record}
				\State $target \gets$ $data(tail)$
				\LineComment{Get the eTLD+1 domain of the target}
				\State $apex \gets$ $eTLDplus1(target)$\label{alg:expired-domains:etldp1-domain}
				\If{$apex$ is registrable}\label{alg:expired-domains:is-registrable}
					\State $V_d \gets V_d \cup domain(chain)$\label{alg:expired-domains:is-vulnerable}
				\EndIf
			\EndIf
		\EndFor
	\EndProcedure
	\end{algorithmic}
\end{algorithm}

\subsection{Discontinued Services}
The process of finding discontinued services is initially performed by traversing each resolving chain to identify whether they point to one of the services supported by our framework. This step is implemented differently according to the documentation provided by each service and typically involves checking for the presence of an \texttt{A} record resolving to a specific IP address, the canonical name of a \texttt{CNAME} record matching a given host or the existence of a \texttt{NS} record pointing to the DNS server of a service.

After completing this stage, (sub)domains mapped to services are checked to verify if the binding between a user account and the domain is in place on the service provider. For the majority of the services considered in this study, this can be achieved by sending a HTTP request to the target domain. The response to this request typically contains an error message that exposes the lack of a correct association of the domain to an account, causing the domains to be vulnerable to a takeover, should an attacker connect it to their account. Other services do not show meaningful error messages and require to actively probing the availability of the domain for a fresh test account that we created on these services. In this case, the process of associating a custom domain to the service has been done using the automated browser testing library puppeteer with Google Chrome~\cite{Puppeteer}.

An improvement over the previous attack involves the detection of DNS wildcards, or -- in specific cases -- the presence of certain DNS records, that might expose vulnerable subdomains as described in §\ref{subsec:abusing-related-domains}. A DNS wildcard for a domain such as \domain{test.example.com} can be easily detected by attempting to resolve a \texttt{CNAME} or \texttt{A} DNS record for \domain{<nonce>.test.example.com}, where \texttt{nonce} refers to a random string that is unlikely to match an entry in the DNS zone of the target domain. The whole high-level procedure adopted by RDScan to discover discontinued services, is summarized in Algorithm~\ref{alg:discontinued-services}.

\begin{algorithm}
	\small
	\caption{Detection of Discontinued Services}\label{alg:discontinued-services}
	\begin{algorithmic}[1]
	\Require{Set of DNS resolving chains $RC$, set of supported services $S$}
	\Ensure{Set of vulnerable subdomains $V_s$}
	\Procedure{discontinued\_services}{$RC, S$}
		\State $V_s \gets \emptyset$
		\For{\textbf{each} $chain \in RC$}
			\For{\textbf{each} $service \in S$}
				\LineComment{Check if a record in the chain points to the service}
				\If{$chain$ points to $service$}\label{alg:discontinued-services:point-to-service}
					\State $d \gets domain(chain)$
					\If{$d$ is available at $service$}\label{alg:discontinued-services:is-available}
						\State $V_s \gets V_s \cup d$
					\Else{ \textbf{if} $service$ supports wildcards \textbf{then}}
						\LineComment{Do wildcard detection if service allows wildcards}
						\State $r \gets generate\_nonce()$
						\State $rd\_chains \gets compute\_resolving\_chains(r.d)$
						\For{\textbf{each} $rd\_chain \in rd\_chains$}
							\If{$rd\_chain$ points to $service$}\label{alg:discontinued-services:wildcard-point-to-service}
								\State $V_s \gets V_s \cup r.d$
							\EndIf
						\EndFor
					\EndIf
				\EndIf
			\EndFor
		\EndFor
	\EndProcedure
	\end{algorithmic}
\end{algorithm}

\subsection{Deprovisioned Cloud Instances}\label{subsubsec:cloud-instances}
The detection of potentially deprovisioned cloud instances has been performed similarly to the probabilistic approach adopted by~\cite{LiuHW16,BorgolteFHKV18} and summarized in Algorithm~\ref{alg:cloud-instances}. We did not create any virtual machine or registered any service at cloud providers in this process. Instead, we collected the set of IP ranges of \dataCloudNProviders{} major providers: Amazon AWS, Google Cloud Platform, Microsoft Azure, Hetzner Cloud, Linode, OVHcloud. We tested each (sub)domain in our dataset the check whether the pointed IP was included in any of the cloud IP ranges (line~\ref{alg:cloud-instances:point-to-cloud}). In case the IP falls within the address range of a cloud provider, we make sure that it does not point to a reserved resource such as a proxy or a load balancer (line~\ref{alg:cloud-instances:not-proxy}).
As a last step, we perform a liveness probe to determine if the IP is in use (line~\ref{alg:cloud-instances:liveness}). This is done by executing a ping to the IP. If no answer is received, we use a publicly available dataset~\cite{Rapid7} comprising a scan of the full IPv4 range on \dataPortscanAll{} ports (\dataPortscanTCP{} TCP, \dataPortscanUDP{} UDP). If no open ports for the given IP are found, we deem the resource as potentially deprovisioned.

\begin{algorithm}
	\small
	\caption{Detection of Deprovisioned Cloud Instances}\label{alg:cloud-instances}
	\begin{algorithmic}[1]
	\Require{Set of DNS resolving chains $RC$, set of cloud providers $C$}
	\Ensure{Set of potentially vulnerable subdomains $V_c$}
	\Procedure{deprovisioned\_cloud}{$RC, C$}
		\State $V_c \gets \emptyset$
		\For{\textbf{each} $chain \in RC$}
			\For{\textbf{each} $\langle provider, ip\_ranges \rangle \in C$}
				\LineComment{Check if a record in the chain points to a cloud provider}
				\If{$chain$ points to $ip \in ip\_ranges$}\label{alg:cloud-instances:point-to-cloud}
					\LineComment{Exclude load balancers and proxies of the $provider$}
					\If{ \textbf{not} $chain$ points to reserved resource}\label{alg:cloud-instances:not-proxy}
						\LineComment{Liveness probe}\label{alg:cloud-instances:liveness}
						\If{ \textbf{not} ( ping($ip$) $\lor$ $ip$ has open ports )}\label{alg:cloud-instances:not-ports}
							\State $V_c \gets V_c \cup domain(chain)$
						\EndIf
					\EndIf
				\EndIf
			\EndFor
		\EndFor
	\EndProcedure
	\end{algorithmic}
\end{algorithm}

\section{Analysis of Discontinued Services}
\label{appendix:services}
Table~\ref{tab:service-capabilities} reports the results of our analysis of third-party services and hosting providers. For each service, we show\begin{enumerate*}[label=]\item the capabilities that are available to its users (Capabilities); \item whether users can claim subdomains of already mapped websites (Wildcard); \item whether the service executes an automatic redirection from the \texttt{www} subdomain of the mapped website and if it can be claimed by others (Redirect); \item the inclusion in the PSL of the parent of the assigned domain if the service allows users to host websites under a specific subdomain (PSL).\end{enumerate*}

The capabilities obtained by an attacker controlling a service vary depending on the specific platform, ranging from content only (UptimeRobot), to full host control (ngrok). In general 19 out of 26 services grant the \capab{js} and \capab{https} capabilities, while 21 grant the \capab{js} capability alone.

The table highlight that the majority of the analyzed services have issues in both the domain ownership verification mechanism and the adoption of the PSL. In particular, only 4 services correctly disallow users to claim subdomains of already mapped domains. Additionally, only 2 of the services which execute the automatic redirect of the \texttt{www} subdomain prevent its mapping from other accounts. Only 2 website of the 20 which offer users a subdomain are present in the PSL. In all other services, registered users control a website which is in a related domain of all other user websites. Moreover, 11 out of 18 services share their primary domain with user controlled websites, which are potential {\rda}s for every other user interacting with the service admin pages.

\begin{table}[ht]
  \footnotesize
  \caption{Attackers' capabilities on vulnerable services.}
  \label{tab:service-capabilities}
  \begin{tabularx}{\linewidth}{@{}lcccX@{}}
    \toprule
    Service & Wildcard & Redirect (\texttt{www}) & PSL & Capabilities \\ \midrule
    agilecrm & \vuln & \notapply & \vuln & \capab{js} \capab{https} \\
    anima  & \vuln & \notapply & \notapply & \capab{js} \capab{https} \\
    campaignmonitor  & \vuln & \notapply & \vuln & \capab{content}* \\
    cargo & \vuln & \safe & \vuln & \capab{js} \\
    feedpress & \vuln & \notapply & \notapply & \capab{html} \\
    gemfury  & \vuln & \notapply & \notapply & \capab{file}$^p$ \capab{https} \\
    github  & \vuln & \notapply & \safe & \capab{js} \capab{file} \capab{https} \\
    helpscout & \vuln & \notapply & \vuln & \capab{js} \capab{file}$^a$ \capab{https} \\
    jetbrains & \safe & \notapply & \vuln & \capab{content}* \\
    launchrock & \vuln & ~~\vuln* & \vuln & \capab{js} \capab{https} \\
    ngrok  & \boh & \boh & \safe & \capab{js} \capab{file} \capab{headers} \capab{https} \\
    persona & \vuln & \safe & \vuln & \capab{js} \capab{https} \\
    pingdom & \vuln & \notapply & \notapply & \capab{js} \\
    readme.io  & \vuln  & \notapply & \vuln & \capab{js} \capab{https} \\
    shopify  & \vuln & \vuln & \vuln & \capab{js} \capab{https} \\
    smartjobboard  & \vuln & \safe & \vuln & \capab{js} \capab{https} \\
    statuspage & \safe & \notapply & \vuln & \capab{js} \capab{https} \\
    strikingly &  \boh &  \boh & \vuln & \capab{js} \capab{https}  \\
    surgesh  & \safe & \safe & \vuln & \capab{js} \capab{https} \\
    tumblr   & \vuln & \notapply & \vuln & \capab{js} \capab{file} \capab{https} \\
    uberflip & \boh & \boh & \notapply & \capab{js} \capab{https} \\
    uptimerobot & \vuln & \notapply & \notapply & \capab{content} \\
    uservoice & \boh & \boh & \vuln & \capab{js} \capab{https} \\
    webflow   & \boh & \boh & \vuln &  \capab{js} \capab{https} \\
    wordpress & \safe & \safe & \vuln & \capab{js} \capab{https} \\
    worksites  & \vuln & \safe  & \vuln & \capab{js} \capab{https} \\
    \hline
  \end{tabularx}
  \ \\ \ 
  {\footnotesize \textbf{Notes:}
    \capab{file}$^a$: only hosts arbitrary \textit{active content} files (js, css);
    \capab{file}$^p$: only hosts arbitrary \textit{passive content} files (images, media, ...);
    \capab{content}*: The altered content is behind an authentication step;
    \vuln*: The launchrock website implicitly associates every subdomain to the mapped domain, not only the \texttt{www} subdomain.
  }
\end{table}

The adoption of the PSL across different dynamic DNS providers is shown in Table~\ref{tab:dyndns-psl}, together with the number of domains that a user can choose from.

\begin{table}[ht]
  \footnotesize
  \caption{PSL on dynamic DNS services.}
  \label{tab:dyndns-psl}
  \begin{tabularx}{\linewidth}{@{}XXX@{}}
    \toprule
    Service & \# Domains & PSL
    \\ \midrule
    afraid (FreeDNS)  & 52,443 & \vuln~0/52,443
    \\
    duckdns & 1& \safe~1/1 \\
    dyndns  & 293 & \vuln~287/293   \\
    noip  & 91 &  \vuln~85/91  \\
    securepoint & 10 & \safe~10/10 \\
    \bottomrule
  \end{tabularx}
\end{table}

\section{Disclosure and Ethical Considerations}
\label{appendix:disclosure}
RDScan identified \dataVSubdomains{} vulnerable subdomains on \dataVDomains{} distinct domains, of which \dataVdSubdomains{} are subdomains pointing to an expired domain and \dataVsSubdomains{} are those mapped to a discontinued service. These domains are the union of the sets $V_d$ and $V_s$ computed as explained in Appendix~\ref{appendix:rdscan}.

We attempted to notify all the websites affected by the issues we discovered. Unfortunately, disclosing the vulnerability information on a large-scale is a challenging task: prior work~\cite{LiDCKB16, StockPRJB16, StockPLBR18} showed that the identification and selection of correct security contact points is the main issue behind an overall unsatisfactory remediation rate. To maximize the chances of a successful notification campaign, we extracted contacts from multiple sources:

\begin{itemize}
    \item \textbf{Vulnerability disclosure programs}\hspace{5pt} Bugcrowd~\cite{Bugcrowd} maintains a comprehensive list of bug bounty and security disclosure programs with detailed contact information. We performed a similarity-based match between each vulnerable domain and the names of the listed programs. We manually verified the correctness of the match to avoid the risk of inadvertently disclose a report for the wrong domain.
    \item \textbf{Security.txt}\hspace{5pt} A recently proposed standard~\cite{SecurityTXT} aims at facilitating the identification of the security contact of a website by including the contact information in the \texttt{security.txt} file placed either in the root directory or under the \texttt{/.well-known/} folder~\cite{RFC8615}. We extracted the security contacts from all the vulnerable websites supporting this practice.
    \item \textbf{Abuse contact database}\hspace{5pt} Abusix provides free access to a database of abuse contact email addresses~\cite{Abusix}. We queried the database with the ip addresses of our set of vulnerable domains and collected the returned email contacts. We filtered out potentially unrelated contacts by discarding those having the domain part of the email address that does not match any of the input domains.
    \item \textbf{Whois data}\hspace{5pt} We also performed WHOIS lookups~\cite{RFC3912} to extract possible contacts. Due to our queries being rate-limited, we used 10 virtual machines running on Google Cloud to lookup all vulnerable domains. The obtained email addresses have been filtered similarly to the Abusix database.
\end{itemize}

We directly disclosed the security issues to all the vulnerable domains having at least one of the aforementioned sources returning a security contact. When multiple contacts have been found for a single website, we gave higher priority to the sources at the top of the list, e.g., we preferred to inform websites over their security disclosure programs than via the email addresses retrieved from WHOIS. Unfortunately, the majority of the considered domains (62\%) did not return any security contact. To inform these websites about their security vulnerabilities, we contacted our national CERT that willingly accepted to disclose the issues to the affected parties on our behalf. Among the few contacted websites with a bug bounty program, F-Secure awarded us with \euro{}250 for the reported subdomain takeover vulnerability.

Aside from vulnerability disclosure programs, our notification campaign is fully automatic: we sent an email to all the identified contacts containing a high-level description of the vulnerabilities and a link to the security advisory on our web application. The advisory presents a detailed description of the problems found for a given domain and a list of all subdomains enumerated by our framework that includes potentially deprovisioned cloud instances and third-party services in use by the website. Our web application also shows the required actions to fix the reported vulnerabilities, a description of our scanning methodology and offers the possibility to opt out from future scans.

\paragraph{Ethical Considerations.} We consciously designed our vulnerability scanning framework to avoid raising network alerts or causing harm to the analyzed targets. Specifically, the subdomain takeover assessment phase has been carried out mostly by DNS queries and simple HTTP requests. Active websites have never been affected by our tests, since we restricted the analysis to abandoned DNS records. We did not perform any large-scale portscan, instead we opted for a public dataset consisting of a scan of the full IPv4 range on \dataPortscanAll{} ports. We also avoided to check the availability of IP addresses on cloud providers by iterating over the creation of multiple virtual machines, since this practice could interfere with the normal operations of the cloud platforms. Similarly, the web analysis module did not execute any attack against the targets, but limited its operations to the passive collection of data (cookies and security policies), simple HTTP requests and client-side testing. Overall, our approach proved to be lightweight and unobtrusive: we did not receive requests from the analyzed websites to opt-out from future scans and no complaints concerning our activity arrived to the abuse contact of the IPs used to perform the analysis.

\end{document}